\documentclass[a4paper,11pt]{article}
\pdfoutput=1 

\usepackage{jcappub} 

\usepackage[normalem]{ulem}
\usepackage[utf8]{inputenc}

\newcommand{\hmpc}{\, \text{Mpc}\,h^{-1}}
\newcommand{\hgpc}{\, \text{Gpc}\,h^{-1}}

\newcommand{\T}{\mathbf{\hat{\mathcal{T}}}}

\newcommand{\Dk}[1]{\frac{d^3#1}{(2\pi)^3}}

\newcommand{\ve}[1]{{\text{\bf #1}}} 
\newcommand{\vk}{\ve k}
\newcommand{\vp}{\ve p}
\newcommand{\vq}{\ve q}
\newcommand{\vx}{\ve x}

\newcommand{\mA}{\mathcal{A}}
\newcommand{\mB}{\mathcal{B}}

\newcommand{\ikk}{\underset{\vk_{12}= \vk}{\int}}
\newcommand{\ikkk}{\underset{\vk_{123}= \vk}{\int}}

\newcommand{\dD}{\delta_\text{D}}

\newcommand{\vhn}{\hat{\ve n}}

\title{\boldmath A Lagrangian Perturbation Theory in the presence of massive neutrinos}

\author[a,b]{Alejandro Aviles}
\emailAdd{avilescervantes@gmail.com}

\author[c,d,e]{Arka Banerjee}
\emailAdd{arkab@stanford.edu}

\affiliation[a]{Consejo Nacional de Ciencia y Tecnolog\'ia, Av. Insurgentes Sur 1582,
Colonia Cr\'edito Constructor, Del. Benito Jurez, 03940, Ciudad de M\'exico, M\'exico}
\affiliation[b]{Departamento de F\'isica, Instituto Nacional de Investigaciones Nucleares,
Apartado Postal 18-1027, Col. Escand\'on, Ciudad de M\'exico,11801, M\'exico.}
\affiliation[c]{Kavli Institute for Particle Astrophysics and Cosmology, Stanford University, 452 Lomita Mall, Stanford, CA 94305, USA}
\affiliation[d]{Department of Physics, Stanford University, 382 Via Pueblo Mall, Stanford, CA 94305, USA}
\affiliation[e]{SLAC National Accelerator Laboratory, 2575 Sand Hill Road, Menlo Park, CA  94025, USA}

\abstract{
We develop a Lagrangian Perturbation Theory (LPT) framework to study the clustering of cold dark matter (CDM) in cosmologies with massive neutrinos. We follow the trajectories of 
CDM particles with Lagrangian displacements fields up to third order in perturbation theory.  Once the neutrinos become non-relativistic, their density fluctuations are modeled as being proportional to the CDM density fluctuations, with a scale-dependent proportionality factor. This yields a gravitational back-reaction that introduces additional scales to the linear growth function, which is accounted for in the higher order LPT kernels. Through 
non-linear mappings from Eulerian to Lagrangian frames, we ensure that our theory has a well behaved large scale behavior free of unwanted UV  divergences, which are common when neutrino and CDM densities are not treated on an equal footing, and 
in resummation schemes that manifestly break Galilean invariance. 
We use our theory to construct correlation functions for both the underlying matter field, as well as for biased tracers using Convolution-LPT. Redshift-space distortions effects are modeled using the Gaussian Streaming Model. When comparing our analytical results to simulated data from the \textsc{Quijote}\footnote{https://github.com/franciscovillaescusa/Quijote-simulations} simulation suite, we
find good accuracy down to $r=20 \hmpc$ at redshift $z=0.5$, for the real space and redshift space monopole particle correlation functions with no free parameters. The same accuracy is reached for the redshift space quadrupole if we additionally consider an effective field theory 
parameter that shifts the pairwise velocity dispersion. For modeling the correlation functions of tracers we adopt a simple Lagrangian biasing scheme with only density and curvature operators, which we find sufficient to 
reach down to $r=20 \hmpc$ when comparing to simulated halos.
}

\begin{document} 
\maketitle
\flushbottom

%
%
%
%
%

\begin{section}{Introduction}\label{sect:Introduction}


Relic neutrinos produced in the early Universe are the second most abundant standard model particles. Hence, despite their 
tiny masses, their contribution to the total cosmological density budget at low redshifts is non-negligible. Neutrino oscillation experiments 
give lower bounds for the sum of their masses, being $~0.06$ eV for a normal hierarchy and $~0.11$ eV for inverted hierarchy \cite{Esteban:2018azc}.  
On the other hand measurements of the Cosmic Microwave Background (CMB) anisotropies from the Planck satellite yield an upper bound of $~0.24$ eV. This tightens to  $~0.12$ eV when combined with
BAO observations \cite{Aghanim:2018eyx}, with strong degeneracies with $H_0$ and $\sigma_8$. 
While the lower mass bounds on neutrino mass imply that massive neutrinos relevant for structure formation must be non-relativistic at late times, 
the thermal velocities of the neutrinos are still relevant down to $z=0$, since they decoupled from the primordial plasma while still relativistic. 
The thermal velocity introduces a new scale in the problem - the free streaming scale \cite{Lesgourgues:2006nd}. 
On scales larger than the free-streaming scale, the neutrino thermal velocity is not large enough to prevent gravitational collapse into the potential wells set by CDM and baryons, 
but on smaller scales, the thermal velocities prevent the growth of the neutrino perturbations. For realistic neutrino masses, this scale is $\sim 100\hmpc$. The other relevant scale is the maximum value of the 
free streaming scale over the history of the Universe's evolution. On scales larger than this, neutrinos, once they are non-relativistic, and CDM behave exactly the same. 
On scales smaller than this scale, the neutrino power spectrum is damped with respect to the CDM power spectrum. For realistic neutrino masses, this scale is $\sim 1 \hgpc$.
Since the growth of the neutrino perturbations on small scales is prevented by the presence of large thermal velocities, the total matter power spectrum, which includes terms arising 
from the CDM-neutrino cross spectrum and the neutrino power spectrum, are damped compared to a massless neutrino cosmology. The CDM power spectrum is itself also damped compared to a massless neutrino cosmology. 
This is due to the fact that the neutrinos contribute to
the source term of the Poisson equation that drives the growth of the CDM perturbations. 
The damping of both the CDM power spectrum, and the total matter power spectrum scale with the neutrino mass, with a different prefactor for each \cite{Lesgourgues:2006nd}.

Ongoing and future galaxy surveys, such as eBOSS\footnote{https://www.sdss.org/surveys/eboss/}, DESI\footnote{https://www.desi.lbl.gov/}, Euclid\footnote{https://sci.esa.int/web/euclid} and LSST\footnote{https://www.lsst.org/}, 
will impose tighter constraints on the sum of the neutrino masses and potentially detect their mass ordering \cite{Hu:1997mj,2009arXiv0912.0201L,Font-Ribera:2013rwa,2013JCAP...01..026A,Hannestad:2016fog}.
As the redshift depth and angular size of galaxy surveys increase, they cover scales where quasi-linear effects 
are more relevant and the tools of Perturbation Theory (PT) become even more important \cite{Bernardeau:2001qr}. 
To fully exploit the forthcoming wealth of data using analytical and semi-analytical methods, comprehensive theories of clustering, valid on these quasi-linear scales, are needed. 
The construction of such models has been widely developed within the massless neutrinos LCDM. 
However, for cosmologies that include the effects of massive neutrinos the situation is rather different, 
and comparatively less work on the subject has been produced so far. Moreover, almost all studies in PT beyond linear order have focused on Fourier space
\cite{Saito:2008bp,Wong:2008ws,Saito:2009ah,Lesgourgues:2009am,Shoji:2009gg,Upadhye:2013ndm,Dupuy:2013jaa,Blas:2014hya,Fuhrer:2014zka,Castorina:2015bma,Levi:2016tlf,Senatore:2017hyk}, 
so non-linear analytical tools for computing the real and redshift space correlation functions are 
still lacking in the literature; but see, e.g., \cite{Peloso:2015jua}, where the authors study the degradation and shift of the BAO peak by using the resummation theory of \cite{Matsubara:2007wj} with Einstein-de Sitter (EdS) kernels.
In \cite{Wright:2017dkw} a 2LPT theory that treats neutrinos as linear is constructed for using it in a hybrid $N$-body/PT, \textsc{COLA} scheme. 

On the other hand, advances in simulation methods have ensured that the effects of massive neutrinos can now be included in $N$-body simulations of structure formation, at the level of accuracy needed for the future surveys. 
While numerous different techniques have been adopted for this purpose \cite{Brandbyge_2009, Viel_2010,Ali_Ha_moud_2012,Banerjee:2016zaa,Inman_2017,Banerjee_2018,hybrid,Dakin_2019}, 
their results mostly agree on the quasi-linear scales of interest in this paper. These simulations, therefore, provide an ideal test-bed for calibrating analytic and semi-analytic models of real-space clustering. 
Once the latter has been tested against measurements from simulations, they can be used over a wide range of cosmological parameter space without having to run computationally expensive simulations.

In this work we construct a Lagrangian Perturbation Theory (LPT) for cold dark matter clustering in the presence of massive neutrinos, accounting for the additional scale dependence introduced by the free-streaming. 
The presence of non-negligible thermal velocities of neutrinos is an added complication for the Lagrangian approach to structure formation in cosmology. 
For CDM particles, all particles starting at initial coordinate $\vq$ follow the same trajectory $\vx(t)$ as a function of time, and coherent flows are found for patches of a few $\text{Mpc}$. 
This leads to a sensible definition of the displacement field, which is the object of interest for Lagrangian approaches. 
Neutrino particles, on the other hand, have very different trajectories as a function of time, even if they start at the same position, due to differences in the magnitude and direction of their thermal 
velocities.\footnote{This complication is also present in the Eulerian approach since
it is not possible to have a well defined velocity field at small scales, and needs to be addressed directly with the Boltzmann equation, as in \cite{Senatore:2017hyk}. 
Other works approximate the neutrinos as a perfect fluid with a Jeans-like mechanism with Jeans length settled by the free-streaming scale \cite{Shoji:2010hm,Blas:2014hya}. } 
This leads to difficulties in defining a single-valued displacement field which captures the correct evolution of neutrino particles ---in \cite{Aviles:2015osc,Cusin:2016zvu} this problem is overcome for the simpler case of a single 
warm dark matter fluid. 
For this reason in this work Lagrangian displacements will follow the trajectories of CDM particles, and non-linear neutrinos overdensities, once non-relativistic, will be approximated 
to be proportional to the CDM fluctuations, where both are equal at scales much larger than the free-streaming, while damped by a factor equal to the ratio of their linear densities at smaller scales. Since we will assume
adiabatic perturbations throughout, this factor is given by the ratio of the transfer functions of neutrinos and CDM.
This approach is analogous to some studies in the Eulerian framework \cite{Lesgourgues:2009am,Upadhye:2013ndm}, where it is shown that this approximation introduces an error of $\sim 0.1\%$.         

With the Lagrangian displacement kernels at hand, we construct the 2-point statistics for the density and velocity fields to 1-loop order in PT. We do this to obtain the real space correlation function 
using the formalism of Convolution-LPT (CLPT) \cite{Carlson:2012bu}, and the redshift space correlation function using the Gaussian Streaming Model (GSM) \cite{Reid:2011ar,Wang:2013hwa}, with the pairwise velocity and
velocity dispersion computed with CLPT. 
Since the presence of the free-streaming introduces further scale-dependencies into the LPT kernels, the way to obtain the ingredients of the GSM differs from the massless neutrino $\Lambda$CDM model; for this endeavour we will
use the formalism recently developed in \cite{Valogiannis:2019nfz} in the context of modified gravity.

We test our theory against data obtained from the \textsc{Quijote} simulations suite \cite{Villaescusa-Navarro:2019bje} for neutrinos with total mass $M_\nu = \sum_i m_{\nu,i} = 0,\,0.1,\, 0.2\,\, \text{and}
\,\, 0.4 \,\text{eV}$ at redshift $z=0.5$, where the masses are distributed equally among the three mass eigenstates.
We find that our formalism is capable of accurately fitting the simulation measurements (for particles) of the real space and redshift-space monopole correlation functions down to $20 \hmpc$ with no free parameters, while the redshift-space quadrupole 
shows the same level of accuracy only if we add an additional Effective Field Theory (EFT) parameter that serves to shift significantly the pairwise velocity dispersion \cite{Reid:2011ar,Vlah:2016bcl}.  
For halos with masses $13.1 < \log_{10}\big[M_h/(M_\odot\,h^{-1})\big] <13.5$, we are capable to reach down to $20 \hmpc$ with the use of linear and second order Lagrangian local biases, and a curvature bias. However, we note that the latter is 
only necessary when the bias is defined with respect to the total correlation function  (i.e. including both CDM and neutrino components), while being consistent with zero when the bias is defined with respect to the CDM correlation function (with the exception of the case $M_\nu=0.4$ eV). 
This agrees with previous studies that show that linear bias, although scale-dependent in 
the presence of massive neutrinos, can be well approximated by a (time-dependent) constant when the biasing prescription is applied to the CDM component only, but not when it is applied to the total matter field
\cite{Villaescusa-Navarro:2013pva,Castorina:2013wga,LoVerde:2014pxa,Vagnozzi:2018pwo,Banerjee:2019omr}.

 The rest of this work is organized as follows. The general formalism is presented in section \ref{sect:EoM}, arriving at the evolution equation for the CDM Lagrangian displacement field in eq.~\eqref{eqm}. In section
 \ref{sect:PT} we find the kernels of the Lagrangian displacement up to third order in PT, given in eqs.~\eqref{KernelPsi1}, \eqref{KernelPsi2} and \eqref{LPTK3order}. 
 In section \ref{sect:deltanu} we discuss our approximation for the neutrino density field and how this
 leads to a well behaved theory free of UV divergences. We construct the real space and redshift space correlation functions in sections \ref{sect:RSCF} and \ref{sect:zSCF} 
 where we also compare to CDM and CDM + neutrino particle simulated data. In section  \ref{sect:halos} we test our formalism against CDM halos. Finally in section \ref{sect:conclusions} we present our conclusions. 
 Some calculations are presented in Appendix \ref{app:kq}.

\end{section}

\begin{section}{Lagrangian displacements}\label{sect:EoM}

The trajectories $\vx(t)$ of cold dark matter particles are related to their initial, Lagrangian positions $\vq$ as
\begin{equation}\label{qTox}
 \vx(t) = \vq + \Psi(\vq,t),
\end{equation}
where $\Psi(\vq,t)$ is the Lagrangian displacement vector field, assumed longitudinal and initially Gaussian distributed. Henceforth, we will omit the time argument, and assume implicitly that they are functions of time.  Using mass conservation one can relate the Lagrangian displacement to the overdensity
\begin{equation}\label{Psitodelta}
  \delta_{cb}(\vx) = \frac{1-J(\vq)}{J(\vq)}.
\end{equation}
Following a standard notation in the literature, subscript ``$cb$'' means that we are referring to the combined CDM-baryons fluid, although we will treat the baryons as CDM particles. $J_{ij}(\vq)=\delta_{ij} + \Psi_{i,j}(\vq)$ is the 
Jacobian matrix of the coordinates transformation (\ref{qTox}) and $J$ its determinant.
The geodesic equation yields
\begin{equation}\label{geodesicEq}
 \nabla_{\vx} \cdot \T \Psi(\vq)  = -\frac{1}{a^2}\nabla^2_{\vx} \Phi(\vx), 
\end{equation}
with $\Phi$ the Newtonian potential. We use $\nabla_\vx = \partial/\partial \vx$ to denote partial derivatives with respect to
Eulerian coordinates. A comma is used to denote differentiation with respect to Lagrangian coordinates. We further define the linear operator \cite{Matsubara:2015ipa}
\begin{equation} 
 \T = \frac{d^2\,}{dt^2} + 2 H \frac{d\,}{dt}, 
\end{equation}
and for notational compactness we introduce
\begin{equation}\label{A0}
 A_0 = 4\pi G \bar{\rho}_m.
\end{equation} 
In ($q$-)Fourier space
\begin{equation}\label{eom}
 \big[\nabla_{\vx} \cdot \T \Psi(\vq)](\vk) = -A_0 f_{cb} \tilde{\delta}_{cb}(\vk) - \tilde{S}(\vk), 
\end{equation}
with $f_{cb} \equiv \Omega_{cb}/\Omega_m$,
and $[(\cdots)](\vk)$ indicates the Fourier transform of $(\cdots)(\vq)$. We have used the Poisson equation 
\begin{equation}\label{Poisson}
 \frac{1}{a^2}\nabla^2_\vx \Phi(\vx) = A_0  f_{cb}  \tilde{\delta}_{cb}(\vk) + \tilde{S}(\vk).  
\end{equation}
In general, we do not write a tilde over Fourier transforms, but we do it in $ \tilde{\delta}_{cb}(\vk)$ and  $ \tilde{S}(\vk)$ to emphasize that they are $q$-Fourier transform
of their Eulerian position counterparts; that is, for a general function $f(\vx)$, 
\begin{equation}\label{tildef}
\tilde{f}(\vk) = \int \Dk{k} e^{i\vk \cdot \vq} f(\vx). 
\end{equation}
The source term $\tilde{S}(\vk)$ is the neutrino density
\begin{equation}
 \tilde{S}(\vk) = A_0 f_{\nu} \tilde{\delta}_\nu(\vk)  \equiv A_0 f_{\nu} \tilde{\alpha}(\vk) \tilde{\delta}_{cb}(\vk), 
\end{equation}
where $f_\nu \equiv \Omega_\nu/\Omega_m = 1 -f_{cb}$, and in the last equality we have introduced the function $\tilde{\alpha}(\vk)\equiv \tilde{\delta}_\nu/ \tilde{\delta}_{cb} = \tilde{\alpha}(k) $, 
where the angular dependencies $\hat{\vk}$ cancel since they are carried by the primordial initial conditions set down by inflation. 
Later we will discuss more about this function $\alpha$. For the moment we will treat $\tilde{S}$ as a source proportional to the $cb$ density perturbations. Our strategy is to write the rhs of 
eq.~(\ref{eom}) in terms of only Lagrangian coordinates. For the CDM density field, eq.~(\ref{Psitodelta}) implies
\begin{align}\label{Psitodk}
 -\tilde{\delta}_{cb}(\vk) &= \Big[\Psi_{i,i} - \frac{1}{2} \big((\Psi_{i,i})^2 + \Psi_{i,j}\Psi_{j,i}\big) \nonumber\\
               &\quad \qquad + \frac{1}{6}(\Psi_{i,i})^3 + \frac{1}{3}\Psi_{i,j}\Psi_{j,k}\Psi_{k,i} -\frac{1}{2}\Psi_{k,k}\Psi_{i,j}\Psi_{j,i} + \cdots \Big](\vk),
\end{align}
where we stop at cubic powers of the Lagrangian displacement. We need now to write the factor $\tilde{\alpha}(\vk)$ of the source $\tilde{S}$ in Lagrangian coordinates. We expand a general funcion $f(\vx)$ around  $\vx=\vq$,
$f(\vx) = f(\vq+\Psi) = f(\vq) + \Psi_i(\vq) f_{,i}(\vq) + \frac{1}{2}\Psi_i(\vq) \Psi_j(\vq) f_{,ij}(\vq) + \cdots$, and with the use of eq.~\eqref{tildef} this implies the following relation between 
$x$- and $q$-Fourier transforms\footnote{Throughout we use the
shorthand notations
\begin{equation}
\vk_{1\cdots n} =   \vk_1 + \cdots + \vk_n,  
\end{equation}
and
\begin{equation}
 \underset{\vk_{1\cdots n}= \vk}{\int} = \int \frac{d^3k_1 \cdots d^3k_n}{(2\pi)^{3n}} (2\pi)^{3} \dD(\vk_{1\cdots n} - \vk).
\end{equation}
}
\begin{align}\label{ftotf}
 \tilde{f}(\vk) &= f(\vk) + \ikk  h_{i}^{f\tilde{f}}(\vk_1,\vk_2) f(\vk_1) \Psi_i(\vk_2)  \nonumber\\
                &\quad   + \ikkk  h_{ij}^{f\tilde{f}}(\vk_1,\vk_2,\vk_3) f(\vk_1) \Psi_{i}(\vk_2)\Psi_{j}(\vk_3) + \cdots,
\end{align}
and the inverse relation
\begin{align}\label{tftof}
 f(\vk)         &= \tilde{f}(\vk) +  \ikk  h_i^{\tilde{f}f}(\vk_1,\vk_2) \tilde{f}(\vk_1) \Psi_i(\vk_2)    \nonumber\\
                &\quad  
                   + \ikkk  h_{ij}^{\tilde{f}f}(\vk_1,\vk_2,\vk_3) \tilde{f}(\vk_1) \Psi_{i}(\vk_2)\Psi_{j}(\vk_3) +\cdots,
\end{align}
with
\begin{align}
 h_{i}^{f\tilde{f}}(\vk_1,\vk_2) &= -  h_{i}^{\tilde{f}f}(\vk_1,\vk_2) = i k_1^i, \\
 h_{ij}^{f\tilde{f}}(\vk_1,\vk_2,\vk_3) &=  -\frac{1}{2}k_1^i k_1^j, \\
 h_{ij}^{\tilde{f}f}(\vk_1,\vk_2,\vk_3) &=  - h_{ij}^{f\tilde{f}}(\vk_1,\vk_2,\vk_3) + h_{i}^{f\tilde{f}}(\vk_{13},\vk_2)h_{i}^{f\tilde{f}}(\vk_1,\vk_3).
\end{align}
We use eqs.~\eqref{ftotf} and \eqref{tftof} to transform $\tilde{\alpha}$ to $\alpha$, yielding
\begin{align}
&- \frac{1}{A_0 f_\nu}\tilde{S}(\vk) = -\alpha(\vk) \tilde{\delta}_{cb}(\vk) 
-  \ikk  (\alpha(\vk_1)-\alpha(\vk)) h_i^{f\tilde{f}}(\vk_1,\vk_2)\tilde{\delta}_{cb}(\vk_1) \Psi_i(\vk_2)   \nonumber\\
& -\ikkk \Big\{  \alpha(\vk) h_{ij}^{\tilde{f}f}(\vk_1,\vk_2,\vk_3) +
               \alpha(\vk_{13}) h_{i}^{f\tilde{f}}(\vk_{13},\vk_2) h_j^{\tilde{f}f}(\vk_1,\vk_3)    \nonumber\\
&\quad   \qquad\qquad               +\alpha(\vk_1) h_{ij}^{f\tilde{f}}(\vk_1,\vk_2,\vk_3)        \Big\}     \tilde{\delta}_{cb}(\vk_1) \Psi_{i}(\vk_2)\Psi_{j}(\vk_3). 
\end{align}
Now, using eq.~(\ref{Psitodk}) in the $cb$ overdensities inside the convolution integrals of the above equation, we can write
\begin{align}\label{sourceSFL}
- \tilde{S}(\vk) &= - A_0 f_\nu \alpha(\vk) \tilde{\delta}_{cb}(\vk) 
         - \ikk  \mathcal{K}^\text{FL$\Psi$}_{ki}(\vk_1,\vk_2) \Psi_k(\vk_1) \Psi_i(\vk_2)  \nonumber\\
&\quad   - \ikkk  \mathcal{K}^\text{FL$\Psi$}_{kij}(\vk_1,\vk_2,\vk_3) \Psi_k(\vk_1) \Psi_{i}(\vk_2)\Psi_{j}(\vk_3), 
\end{align}
with
\begin{align}
 \mathcal{K}^\text{FL$\Psi$}_{ki}(\vk_1,\vk_2) &=  A_0 f_\nu (\alpha(\vk_1)-\alpha(\vk))  k_1^k k_1^i, \\
 \mathcal{K}^\text{FL$\Psi$}_{kij}(\vk_1,\vk_2,\vk_3) &= -iA_0 f_\nu \big(\alpha(\vk)-\alpha(\vk_1) \big) k_1^k k_1^i k_1^j \nonumber\\
 &\quad    +  i A_0 f_\nu \big( \alpha(\vk) - \alpha(\vk_{13}) \big) (k_1^i+k_3^i) \left[ k_1^jk_1^k +\frac{1}{2} k_1^k k_3^j + \frac{1}{2}k_1^j k_3^k \right].            
\end{align}
We will refer to the  terms that contain the kernels $\mathcal{K}^\text{FL$\Psi$}_{ij\cdots}$ as ``frame-lagging'', since they arise when mapping Fourier transforms between Eulerian and Lagrangian frames. These terms are necessary in 
LPT frameworks beyond $\Lambda$CDM with additional scales defined in Eulerian coordinates. They were used first in \cite{Aviles:2017aor} in the context Modified Gravity (MG), in a different, less general method we have followed here, 
and later in other MG works \cite{Winther:2017jof,Aviles:2018saf,Valogiannis:2019xed,Moretti:2019bob} and for a COLA implementation with MG and massive neutrinos \cite{Wright:2017dkw}.

Now, introducing the function 
\begin{equation}\label{Aofk}
 A(k) = A_0 \big[ f_{cb} + f_\nu \alpha(k) \big],
\end{equation}
eq.~(\ref{eom}) becomes
\begin{align}\label{preLDEoM}
 [\nabla_\vx \cdot \T \Psi](\vk) &= -A(k) \tilde{\delta}_{cb}(\vk)  
 - \ikk  \mathcal{K}^\text{FL$\Psi$}_{ki}(\vk_1,\vk_2) \Psi_k(\vk_1) \Psi_i(\vk_2)  \nonumber\\
           &\quad -\ikkk  \mathcal{K}^\text{FL$\Psi$}_{kij}(\vk_1,\vk_2,\vk_3) \Psi_k(\vk_1) \Psi_{i}(\vk_2)\Psi_{j}(\vk_3). 
\end{align}
Using $\nabla_\vx \cdot \T \Psi = (J^{-1})_{ij} \T \Psi_{i,j} = \T \Psi_{i,i} - \Psi_{i,j}\T \Psi_{i,j} + \Psi_{i,k}\Psi_{k,j}\T \Psi_{i,j} + \cdots$ and
eq.~(\ref{Psitodk}) we arrive at the equation of motion for the displacement field
\begin{align} \label{eqm}
 &\big(\T - A(k) \big) [\Psi_{i,i}](\vk) = [\Psi_{i,j} \T \Psi_{j,i}](\vk) 
 - \frac{A(k)}{2}[\Psi_{i,j} \Psi_{j,i}](\vk) - \frac{A(k)}{2} [(\Psi_{l,l})^2](\vk)  \nonumber\\*
 &\quad - [\Psi_{i,k}\Psi_{k,j} \T \Psi_{j,i}](\vk) +\frac{A(k)}{6} [(\Psi_{l,l})^3](\vk) 
  + \frac{A(k)}{2} [\Psi_{l,l} \Psi_{i,j} \Psi_{j,i}](\vk) + \frac{A(k)}{3} [\Psi_{i,k}\Psi_{k,j} \Psi_{j,i}](\vk) \nonumber\\*
  &\quad 
-  \ikk  \mathcal{K}^\text{FL$\Psi$}_{ki}(\vk_1,\vk_2) \Psi_k(\vk_1) \Psi_i(\vk_2) 
-   \ikkk  \mathcal{K}^\text{FL$\Psi$}_{kij}(\vk_1,\vk_2,\vk_3) \Psi_k(\vk_1) \Psi_{i}(\vk_2)\Psi_{j}(\vk_3),
\end{align}
valid up to cubic powers of $\Psi$, which is sufficient to construct LPT kernels up to third order, as we do in the following section.
Notice that at very large scales both massive neutrinos and $cb$ density perturbations are equal, $\alpha(k)$ becomes 1, and 
\begin{equation}
A(k\rightarrow 0) \rightarrow A_0 = 4 \pi G \bar{\rho}_m, 
\end{equation}
which simply means that neutrinos behave indistinguishably from CDM. 

If $\alpha=1$, the frame-lagging kernels  vanish and we recover the standard equation 
for the longitudinal piece of the Lagrangian displacement in $\Lambda$CDM; see, e.g., \cite{Matsubara:2015ipa}.
Moreover, in that case eq.~\eqref{eqm} becomes exact.

\end{section}

\begin{section}{Perturbation Theory} \label{sect:PT}

In this section we find formal solutions to the Lagrangian displacements up to third order in PT. That is, 
as usual, we expand $\Psi=\Psi^{(1)} + \Psi^{(2)} + \Psi^{(3)} + \cdots$, and solve eq.~\eqref{eqm} in an iterative manner.

To linear order we use eq.~\eqref{Psitodelta} to connect density and Lagrangian displacement linear fields as
%
\begin{equation} \label{Psi1}
 \Psi_i^{(1)}(\vk,t) = i\frac{k_i}{k^2} \delta_{cb}^{(1)}(\vk,t),\qquad \text{with} \qquad \delta_{cb}^{(1)}(\vk,t)= \delta_{cb}^{(1)}(\vk,t_0) D_+(\vk,t), 
\end{equation}
and the scale-dependent linear growth function $D_+$ is the growing solution to 
\begin{equation} \label{DplusEv}
 \big(\T - A(k) \big) D_+(\vk,t) = 0,
\end{equation}
as obtained from eq.~\eqref{eqm}. At linear order, this yields $ (\T - A(k)) [\Psi_{i,i}^{(1)}](\vk) = 0$.
To solve the above equation we start the evolution well inside the matter dominated Universe evolution phase, but once the neutrinos are non-relativistic, and use the 
fitting formula presented in \cite{Hu:1997vi} for the evolution of linear $cb$ growth functions during the EdS epoch as initial conditions to eq.~\eqref{DplusEv}.

In general, the Lagrangian displacement to $n$-th order is
\begin{equation}
\Psi_i^{(n)}(\vk,t) = \frac{i}{n!} \underset{\vk_{1\cdots n}= \vk}{\int} L^{(n)}_i(\vk_1,\cdots,\vk_n;t) D_+(\vk_1,t) \cdots D_+(\vk_n,t) \delta_1 \cdots \delta_n
\end{equation}
where $\delta_1=\delta_{cb}^{(1)}(\vk_1,t_0)$,  $\delta_2=\delta_{cb}^{(1)}(\vk_2,t_0)$, and so on. From eq.~(\ref{Psi1}) we obtain 
\begin{equation} \label{KernelPsi1}
L_i^{(1)}(\vk) = \frac{k^i}{k^2}. 
\end{equation}
Higher order solutions are found solving eq.~(\ref{eqm}) iteratively, as we do below.

\begin{subsection}{2LPT}

Now, we find the second order LPT kernel by inserting the linear solution \eqref{Psi1} into the rhs of eq.~\eqref{eqm}.
To do so, we first express the integrals containing the frame-lagging contributions in terms of linear density fields, which up to second order only appear through $\mathcal{K}^\text{FL}_{ki}$. We obtain  
\begin{equation}
 -  \ikk  \mathcal{K}^\text{FL$\Psi$}_{ki}(\vk_1,\vk_2) \Psi_k^{(1)}(\vk_1) \Psi_i^{(1)}(\vk_2)  
 = -\frac{1}{2} \ikk K^{(2)}_\text{FL}(\vk_1,\vk_2) D_+(\vk_1)D_+(\vk_2) \delta_1\delta_2
\end{equation}
with
\begin{equation}
 K^{(2)}_\text{FL}(\vk_1,\vk_2) = (A(k)-A(k_1) ) \frac{\vk_1\cdot \vk_2}{k_2^2}
     +(A(k)-A(k_2) ) \frac{\vk_1\cdot \vk_2}{k_1^2},
\end{equation}
with $\vk=\vk_1+\vk_2$. 
Hence, to second order in PT, the equation of motion for the Lagrangian displacement is
\begin{align}\label{eqm2}
 \big(\T - A(k)\big) [\Psi_{i,i}^{(2)}](\vk) &= [\Psi_{i,j}^{(1)} \T \Psi_{j,i}^{(1)}](\vk) 
 - \frac{A(k)}{2}[\Psi_{i,i}^{(1)}\Psi_{j,j}^{(1)} + \Psi_{i,j}^{(1)} \Psi_{j,i}^{(1)}](\vk)
   \nonumber\\
 &\quad-\frac{1}{2} \ikk K^{(2)}_\text{FL}(\vk_1,\vk_2) D_+(k_1)D_+(k_2) \delta_1\delta_2.
\end{align} 
The second order kernel becomes
\begin{align}\label{KernelPsi2}
 L^{(2)}_i(\vk_1,\vk_2) =  \frac{3}{7} \frac{k^i}{k^2} \left( \mA(\vk_1,\vk_2) - \mB(\vk_1,\vk_2) \frac{(\vk_1\cdot\vk_2)^2}{k_1^2 k_2^2} \right). 
\end{align}
$\mA$ and $\mB$ are scale and time dependent functions, defined as
\begin{equation}
\mA = \frac{7}{3}\frac{D_\mA(\vk_1,\vk_2)}{D_+(k_1)D_+(k_2)},\qquad  \mB = \frac{7}{3}\frac{D_\mB(\vk_1,\vk_2)}{D_+(k_1)D_+(k_2)},
\end{equation}
with second order growth functions $D_{\mA,\mB}$ the solutions to second order linear differential equations
\begin{align}
\big(\T - A(k)\big) D_\mA(\vk_1,\vk_2) &= \Big[ A(k) + K^{(2)}_\text{FL}(\vk_1,\vk_2) \Big]D_+(k_1)D_+(k_2), \label{DA}\\
\big(\T - A(k)\big) D_\mB(\vk_1,\vk_2) &= \Big[ A(k_1) + A(k_2) - A(k)  \Big] D_+(k_1)D_+(k_2), \label{DB}    
\end{align}
with appropriate initial conditions to project out the homogeneous, linear order solution. 
For $\Lambda$CDM evolution with no massive neutrinos, $A(k) = \frac{3}{2} \Omega_m H^2$, hence $D_\mA=D_\mB$ are only time dependent,
\begin{equation}\label{DAfnu0}
 D_\mA^{f_\nu = 0}(t)= \frac{3}{7}D_+^2(t) +  \frac{4}{7}\left(\T - \frac{3}{2} \Omega_m H^2  \right)^{-1} \left[ \frac{3}{2} \Omega_m H^2 
 \left( 1- \frac{f^2}{\Omega_m}\right) \right],
\end{equation}
with $f=d\log D_+(t)/d\log a(t)$ the logarithmic growth factor.
For EdS, $\Omega_m=1=f$, and the second term in the rhs of the above equation vanishes, reducing the second order kernel [eq.~\eqref{KernelPsi2}] to the well-known EdS result with $\mA=\mB=1$.

It is useful to define the second order growth function $ D^{(2)}$ as
\begin{equation}
 k_i \Psi_i^{(2)} = \frac{i}{2} \ikk D^{(2)}(\vk_1,\vk_2) \delta_1 \delta_2.
\end{equation}
At large scales $\mA=\mB$, as can be deduced by taking the limit $\vk=0$, $\vk_1=-\vk_2 =\vp$ in eqs.~\eqref{DA} and \eqref{DB}. Hence $D^{(2)}(-\vp,\vp) = 0$, as required since
these wave-vector configurations correspond to planar collapse, for which Zeldovich approximation is exact \cite{McQuinn:2015tva}. 
We notice that this was possible because of cancellations provided by the frame-lagging terms, so these are particularly important to obtain a proper convergence at large scales. 
More generally, to leading order in $k$ one gets
\begin{equation} \label{D2limitk0}
  D^{(2)}(\vk-\vp,\vp) =  \frac{3 }{7} \mathcal{C}_2   \big[ D_+^{M_\nu=0}(t) \big]^2 \left(1-  (\hat{\vk}\cdot \hat{\vp})^2  \right) \frac{k^2}{p^2}, 
\end{equation}
for $k \ll p$, where $\mathcal{C}_2$ is a constant of order unity that depends very weakly on time. It also weakly depends on the wave-vector $\vp$ through $D_+(p,t)$, but it stabilizes beyond the free-streaming scale since 
 $D_+(p\gg k_\text{FS},t) \approx \big[ D_+^{M_\nu=0}(t) \big]^{1- 3/5 f_\nu} $ tends to a scale independent function \cite{Hu:1997vi}. Hence, more precisely, eq.~\eqref{D2limitk0} is valid for $k\ll k_\text{FS} \ll p$.
For example, for massless neutrinos cosmologies  $\mathcal{C}_2(z=0.5) \approx 1.005$, being slightly different to unity because of the contribution of the second term on the rhs of eq.~\eqref{DAfnu0}; 
for degenerated massive neutrinos with total mass $M_\nu=0.4$, we obtain  $\mathcal{C}_2(z=0.5) \approx 0.93$. 

\end{subsection}

\begin{subsection}{Third order Lagrangian displacements}

Now, in this subsection we find solutions to eq.~\eqref{eqm} to third order in PT.
We use the first and second order Lagrangian displacements to write the frame-lagging terms to third order as
\begin{align} 
& -\frac{1}{6} \ikk K^{(3)}_\text{FL}(\vk_1,\vk_2,\vk_3) D_+(k_1)D_+(k_2)D_+(k_3) \delta_1\delta_2\delta_3  \nonumber\\
&\equiv -  \ikk  \mathcal{K}^\text{FL$\Psi$}_{ki}(\vk_1,\vk_2) (\Psi_k^{(2)}(\vk_1) \Psi_i^{(1)}(\vk_2) +\Psi_k^{(1)}(\vk_1) \Psi_i^{(2)}(\vk_2)) \nonumber\\
&\quad  -   \ikkk  \mathcal{K}^\text{FL$\Psi$}_{kij}(\vk_1,\vk_2,\vk_3) \Psi_k^{(1)}(\vk_1) \Psi_{i}^{(1)}(\vk_2)\Psi_{j}^{(1)}(\vk_3) 
\end{align}
with
\begin{align} \label{FL3} 
 &K^{(3)}_\text{FL}(\vk_1,\vk_2,\vk_3) = 3 (A(k)-A(k_1)) \left[\frac{\vk_1 \cdot \vk_{23}}{k_{23}^2}  \frac{D^{(2)}(\vk_2,\vk_3)}{D_+(\vk_2)D_+(\vk_3)} 
 -2 \frac{(\vk_1\cdot \vk_2)(\vk_1\cdot \vk_3)}{k_2^2 k_3^2} \right]\nonumber\\
 &\quad + 3 \big( A(k) - A(k_{23}) \big) \frac{\vk_{1}\cdot\vk_{23}}{k_1^2} 
      \left[  1 + 2\frac{(\vk_2\cdot \vk_3)}{k_3^2} + \frac{(\vk_2\cdot \vk_3)^2}{k_2^2k_3^2} + \frac{D^{(2)}(\vk_2,\vk_3)}{D_+(\vk_2)D_+(\vk_3)}\right], 
\end{align}
and $\vk=\vk_1+\vk_2+\vk_3$. Now, the difference between the $\mathcal{K}_{ij\cdots}^{\text{FL}\Psi}$ and $K_{\text{FL}}$ kernels should be more clear, 
the former serve to expand $\big( \tilde{\alpha}-\alpha \big) \tilde{\delta}_{cb}$ on a Fourier series of non-linear Lagrangian displacements, while the latter serve to expand it on linear density fields.

To third order, the Lagrangian displacement equation of motion [eq.~\eqref{eqm}] becomes
\begin{align} \label{eqm3}
 &(\T - A(k)) [\Psi_{i,i}^{(3)}](\vk) = [\Psi_{i,j}^{(2)} \T \Psi_{j,i}^{(1)}](\vk) +[\Psi_{i,j}^{(1)} \T \Psi_{j,i}^{(2)}](\vk) 
                     - A(k)[\Psi_{i,j}^{(2)} \Psi_{j,i}^{(1)} + \Psi_{i,i}^{(1)}\Psi_{j,j}^{(2)}](\vk)   \nonumber\\
 &  - [\Psi_{i,k}^{(1)}\Psi_{k,j}^{(1)} \T \Psi_{j,i}^{(1)}](\vk) + \frac{A(k)}{3} [\Psi_{i,k}^{(1)}\Psi_{k,j}^{(1)} \Psi_{j,i}^{(1)}](\vk) 
 +\frac{A(k)}{6} [\Psi_{i,i}^{(1)}\Psi_{j,j}^{(1)}\Psi_{k,k}^{(1)}](\vk)  \nonumber\\
  &  + \frac{A(k)}{2} [\Psi_{l,l}^{(1)} \Psi_{i,j}^{(1)} \Psi_{j,i}^{(1)}](\vk) 
  -\frac{1}{6} \ikk K^{(3)}_\text{FL}(\vk_1,\vk_2,\vk_3) D_+(\vk_1)D_+(\vk_2)D_+(\vk_3) \delta_1\delta_2\delta_3.
\end{align}
Inserting the solutions for the first and second order Lagrangian displacements, a lengthy computation leads 
to\footnote{For details, we refer the reader to ref.~\cite{Aviles:2017aor}, where an analogous computation is performed in the context of modified gravity.}
 \begin{align} \label{LPTK3order}
  L^{(3)}_i(\vk_1,\vk_2,\vk_3) &=  \frac{k^i}{k^2}
    \Bigg\{ \frac{5}{7} \left( \mathcal{A}^{(3)} -\mB^{(3)}
              \frac{(\vk_2 \cdot \vk_3)^2}{k^2_2 k^3_2} \right) 
             \left( 1-  \frac{(\vk_1 \cdot \vk_{23})^2}{k_1^2 k_{23}^2} \right)  \nonumber\\*
    &  \quad -\frac{1}{3} \left( \, \mathcal{C}^{(3)}  -3 \mathcal{D}^{(3)} \frac{(\vk_2 \cdot \vk_3)^2}{k^2_2 k^2_3} 
     + 2 \mathcal{E}^{(3)} \frac{(\vk_1 \cdot \vk_2)(\vk_2 \cdot \vk_3)(\vk_3 \cdot \vk_1)}{k_1^2 k^2_2 k^2_3} \, \right)
     \Bigg\},                             
 \end{align} 
plus a transverse piece that does not enter in 2-point, 1-loop statistics. The normalized growth functions are 
\begin{align}
 \mathcal{A}^{(3)},\mathcal{B}^{(3)}(\vk_1,\vk_2,\vk_3) 
     &= \frac{7}{5}  \frac{D^{(3)}_{\mathcal{A},\mathcal{B}}(\vk_1,\vk_2,\vk_3)}{D_{+}(k_1)D_{+}(k_2)D_{+}(k_3)}, \\
 \mathcal{C}^{(3)},\mathcal{D}^{(3)},\mathcal{E}^{(3)}(\vk_1,\vk_2,\vk_3) 
     &= \frac{  D^{(3)}_{\mathcal{C},\mathcal{D},\mathcal{E}}(\vk_1,\vk_2,\vk_3)}{D_{+}(k_1)D_{+}(k_2)D_{+}(k_3)},
\end{align} 
and third order growth functions
\begin{align} 
 \big(\T - A(k)\big)D^{(3)}_\mathcal{A} &=   3 D_+(k_1) \big(A(k_1) + \T - A(k)\big)D^{(2)}_\mA(\vk_2,\vk_3)  , \\
 \big(\T - A(k)\big)D^{(3)}_\mathcal{B} &=    3 D_+(k_1) \big(A(k_1) + \T - A(k)\big)D^{(2)}_\mB(\vk_2,\vk_3)  , \\
 \big(\T - A(k)\big)D^{(3)}_\mathcal{C} &=    9 D_+(k_1) \big(A(k_1) + \T - 2 A(k)\big)D^{(2)}_\mA(\vk_2,\vk_3)
                                            -3 A(k) D_+(k_1)D_+(k_2)D_+(k_3)  \nonumber\\
                              &\quad  + 3 K^{(3)}_\text{FL}(\vk_1,\vk_2,\vk_3) D_+(k_1)D_+(k_2)D_+(k_3) \\                          
 \big(\T - A(k)\big)D^{(3)}_\mathcal{D} &=    3 D_+(k_1) \big(A(k_1) + \T - 2 A(k)\big)D^{(2)}_\mB(\vk_2,\vk_3)
                                            +3 A(k) D_+(k_1)D_+(k_2)D_+(k_3)  , \\
 \big(\T - A(k)\big)D^{(3)}_\mathcal{E} &=    3 \big(3 A(k_1) - A(k)\big)D_+(k_1)D_+(k_2)D_+(k_3).
\end{align}  
It is straightforward to check that for EdS evolution one has $\mathcal{A}^{(3)}=\mathcal{B}^{(3)}=
 \mathcal{C}^{(3)}=\mathcal{D}^{(3)}=\mathcal{E}^{(3)}=1$.\footnote{Use the identities 
 $\T D_+^2 = 2 D_+\T D_+ + 2 \dot{D}_+$ and $(\T- \frac{3}{2} H^2)^{-1} [\frac{3}{2}H^2 D_+^3] = \frac{1}{6}D_+^3$, where $D_+$ is the growing solution to $(\T- \frac{3}{2} H^2)D_+=0$, and $H=2/(3t)$.}
 For $\Lambda$CDM with $f_\nu=0$, these functions are only time dependent,  at $z=0$ $\mathcal{A}^{(3)}=\mathcal{B}^{(3)}\simeq 1.02$, 
 $\mathcal{C}^{(3)}=\mathcal{D}^{(3)}=\mathcal{E}^{(3)}\simeq 1.01$ for typical cosmological parameter values.

We define $D^{(3)}(\vk_1,\vk_2,\vk_3) = k_i L_i^{(3)}(\vk_1,\vk_2,\vk_3) D_+(\vk_1) D_+(\vk_2) D_+(\vk_3)$, hence 
\begin{align}
 k_i \Psi_i^{(3)} = \frac{i}{6} \ikk D^{(3)}(\vk_1,\vk_2,\vk_3) \delta_1 \delta_2 \delta_3.
\end{align}

The relevant configurations for computing 2-point statistics  are double squeezed, for which $\vk_1=\vk$ and $\vk_3=-\vk_2=\vp$. 
Symmetrizing the third order kernel, and evaluating in this configuration we obtain
\begin{align} 
 & \left(\T - A(k)\right) D^{(3)s} (\vk,-\vp,\vp)=   \Bigg\{  D_+(p)\left(A(p) + \T - A(k)\right) D^{(2)}(\vp,\vk) 
                \left( 1- \frac{(\vp \cdot (\vk + \vp))^2}{p^2 |\vp+\vk|^2}\right) \nonumber\\* 
&\quad  + \Bigg[ \big(2A(k) -  A(p) - A(|\vk + \vp|) \big) \left( \frac{D^{(2)}(\vp,\vk)}{D_+(k)D_+(p)}  + 1 + \frac{(\vk \cdot \vp)^2}{k^2p^2} \right)  
              \nonumber\\*
&\quad     \qquad \qquad + A(k) - A(p)   - K^{(2)}_\text{FL}(\vp,\vk)  + K^{(3)}_\text{FL}(-\vp,\vp,\vk) \Bigg] D_+(k)D_+^2(p) \,\,\Bigg\}  \nonumber\\*
&  
 + \quad (\, \vp \rightarrow -\vp \,), 
\end{align}
with
\begin{align}
 K^{(3)}_\text{FL}(-\vp,\vp,\vk)&= (A(p)-A(k)) \frac{\vp\cdot (\vk+\vp)}{|\vk+\vp|^2}  \frac{D^{(2)}(\vk,\vp)}{D_+(k)D_+(p)}  \nonumber\\
 &+     \frac{\vp\cdot (\vk+\vp)}{p^2} (A(|\vk+\vp|)-A(k)) \left[\frac{D^{(2)}(\vk,\vp)}{D_+(k)D_+(p)} + 1 + \frac{(\vk\cdot\vp)^2}{k^2 p^2} \right]  \nonumber\\
 &+        \left[ \frac{(k^2+p^2)(\vk\cdot\vp)^2}{k^2p^4} +   \frac{(k^2+p^2)(\vk\cdot\vp)}{k^3p^3}\right] \big( A(|\vk+\vp|)- A(k) \big). 
\end{align}
One can check that, due to  cancellations provided by the frame-lagging terms, the symmetrized $D^{(3)s} (\vk = 0,-\vp,\vp) \rightarrow 0$, or at leading order in 
$k/p$
\begin{equation} \label{D3LS}
  D^{(3)s} (\vk,-\vp,\vp)=  \frac{7 }{15} \mathcal{C}_3   \big[ D_+^{M_\nu=0}(t) \big]^3 \big[1-  (\hat{\vk}\cdot \hat{\vp})^2  \big]^2 \frac{k^2}{p^2},
\end{equation}
with $ \mathcal{C}_3$ a constant of order unity. Hence, as in  the case of the second order growth function, the leading term when $k\ll p$ is of order $k^2/p^2$.

\vspace{0.2cm}

We end this section by noticing that in ref.~\cite{Aviles:2017aor} the LPT kernels for modified gravity theories were obtained using an approach that considers 
the evolution of a scalar field Klein-Gordon like equation. The kernels obtained in that work are a special case of the kernels obtained 
here.\footnote{One can check that for a function
\begin{equation}
 A^\text{MG}(k) = \frac{3}{2}\Omega_m H^2 \left( 1 + \frac{k^2/a^2}{3 \Pi(k,a)}\right),
\end{equation}
with $\Pi(k,a) = (k^2/a^2 + m^2_\text{MG})/6\beta^2$, one recovers the kernels of \cite{Aviles:2017aor}.} Hence, the method developed in this work is more general and find applications for scenarios that have 
additional scales than $\Lambda$CDM, such as modified gravity or dark matter clustering in the presence of massive neutrinos.

\end{subsection}

\end{section}

\begin{section}{Neutrino density}\label{sect:deltanu}

In the presence of massive neutrinos, and at sufficiently late times such that relativistic components can be neglected, the Poisson equation becomes
\begin{equation}
 \nabla^2_\vx \Phi(\vx,t) = 4\pi G a^2 \bar{\rho}_m (f_{cb}\delta_{cb} + f_{\nu}\delta_{\nu}). 
\end{equation}
To our knowledge, a full, consistent analytic treatment of the non-linear nature of the neutrino density field does not exist in the literature. Various approximations have been tested in studies. 
In the pioneering work of \cite{Saito:2008bp} the $\delta_\nu$ contribution is neglected to obtain the non-linear $\delta_{cb}$, 
such that they use the EdS, SPT kernels, but using the linear power spectrum of the $cb$ fluid, $P_{cb}^L(k)$, to compute the loop corrections. Other works approximate the neutrino overdensity by its linear value and use it as an external source to 
compute non-linear CDM overdensities  \cite{Saito:2009ah,Wong:2008ws,Wright:2017dkw}. In \cite{Blas:2014hya} it was noted that this approach violates momentum conservation 
yielding an incorrect behavior at large scales. In particular, SPT kernels do not follow $F_n \propto k^2$, and 
$(P_\text{NL}-P_\text{L})/P_\text{L} \propto k^2$, as $k \rightarrow 0$. The approach of \cite{Blas:2014hya}, instead, evolves non-linear neutrino density fields by truncating the Boltzmann hierarchy at the 
Euler equation, and approximates the second moment of the phase-space distribution function (the velocity dispersion) to be proportional to an effective sound speed times the density contrast, as in \cite{Shoji:2009gg,Shoji:2010hm,Castorina:2015bma} 
(see also Appendix C of \cite{Aviles:2015osc}).  Reference \cite{Senatore:2017hyk} performs non-linear perturbations around a Fermi-Dirac massive neutrino distribution to solve the coupled Boltzmann and CDM 
density field equations iteratively by expanding in powers of $f_\nu$, keeping only the linear terms in $f_\nu$. 

In this work we will approximate 
\begin{equation}
\tilde{\alpha}(k) \equiv  \frac{\tilde{\delta}_\nu}{\tilde{\delta}_{cb}}  \approx  \frac{\tilde{\delta}^{(1)}_\nu}{\tilde{\delta}^{(1)}_{cb}},   
\end{equation}
where we have returned to the Lagrangian treatment of the previous section --- a tilde means the $q$-Fourier transform of Eulerian-coordinates valued function. 
Within this approximation, the non-linear neutrinos fluctuations become 
\begin{align} \label{deltanuProxy}
\tilde{\delta}_\nu(\vk) = \frac{\tilde{\delta}^{(1)}_\nu}{\tilde{\delta}^{(1)}_{cb}} \tilde{\delta}_{cb}(\vk) 
&= \frac{\delta^{(1)}_\nu}{\delta^{(1)}_{cb}}  \tilde{\delta}_{cb}(\vk) + \frac{f_{cb}}{A_0 f_{\nu}} \ikk \mathcal{K}^\text{FL$\Psi$}_{ki}(\vk_1,\vk_2) \Psi_k(\vk_1) \Psi_i(\vk_1) \nonumber\\
&\quad  + \frac{f_{cb}}{A_0 f_{\nu}} \ikk \mathcal{K}^\text{FL$\Psi$}_{kij}(\vk_1,\vk_2,\vk_3) \Psi_k(\vk_1) \Psi_i(\vk_1)\Psi_i(\vk_3) + \cdots
\end{align}
where in the second equality we use eq.~\eqref{sourceSFL} and, by virtue of eq.~\eqref{Psitodk}, the neutrino density becomes written as a function of Lagrangian displacements and the ratio of linear overdensity fields. 
A similar approximation was adopted in \cite{Lesgourgues:2009am,Upadhye:2013ndm,Levi:2016tlf} for Eulerian space, but notice that these are not exactly equal to ours, since 
``tilded'' functions are given by eq.~\eqref{tildef}. Hence they
carry the non-linear evolution provided by the Lagrangian displacements, as is manifest in the second equality of eq.~\eqref{deltanuProxy}, which contains corrections up to third order in PT.  
The authors of \cite{Blas:2014hya} argue that the approximation $\delta_\nu = (\delta_\nu^{(1)}/\delta_{cb}^{(1)}) \delta_{cb}$ also
violates momentum conservation.
However, a good large scale behavior in our approach is provided by the frame-lagging terms, as
was shown  in \cite{Aviles:2017aor} (sect. IV) in the context of MG, and we show in the following.

From the LPT kernels we construct the SPT kernels as \cite{Matsubara:2011ck,Rampf:2012xa,Aviles:2018saf}
\begin{align}
F_2(\vk_1,\vk_2) &= \frac{1}{2} \big( k_iL_i^{(2)}(\vk_1,\vk_2) + k_ik_j L_i^{(1)}(\vk_1)L_j^{(1)}(\vk_2) \big),   \label{LtoF2}\\
F^s_3(\vk_1,\vk_2,\vk_3) &= \frac{1}{3} \big( k_i L_i^{(3)s}(\vk_1,\vk_2,\vk_3) + k_ik_j(L_i^{(2)}(\vk_1,\vk_2)L_j^{(1)}(\vk_3) + \text{cyclic} ) \nonumber\\
                         &\quad + k_i k_j k_k   L_i^{(1)}(\vk_1)L_j^{(1)}(\vk_2) L_k^{(1)}(\vk_3) \big). \label{LtoF3}
\end{align}
Using eq.~\eqref{D2limitk0} we obtain that $F_2(\vk_1,\vk_2)  \propto k^2$ as $k = |\vk_1+\vk_2| \rightarrow 0$ as required by momentum conservation. The case of 
the third order SPT kernel is challenging since the equations to construct $L^{(3)}_i$ are cumbersome for an analytical treatment, but using eq.~\eqref{D3LS} for the particular configuration 
used in constructing the 1-loop power spectrum it follows that $F_3^s(\vk,-\vp,\vp)$ goes as   $k^2$ for $k \ll p$.

 \begin{figure}
 	\begin{center}
 	\includegraphics[width=3 in]{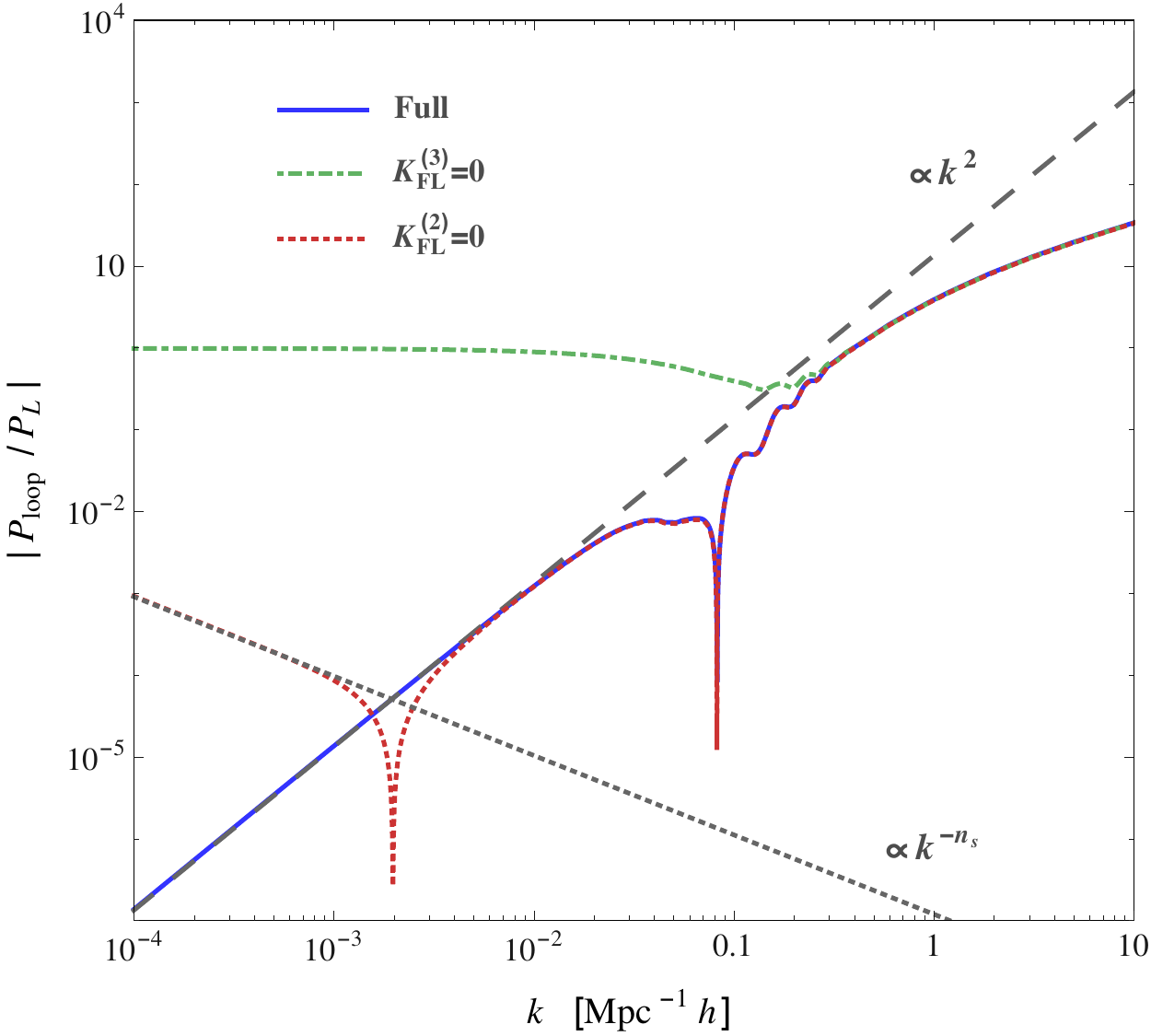}
 	\caption{1-loop correction to the matter power spectrum given by eqs.~\eqref{LtoF2}, \eqref{LtoF3} and \eqref{PcbcbSPT} for massive neutrinos with total mass $M_\nu=0.4\,\text{eV}$. 
 	We plot their ratio to the linear power spectrum showing that at large scales $(P^\text{NL}- P^L)/P^L \propto k^2$.  
 	The blue line shows the full power spectrum including the frame-lagging contributions.
 	Dot-dashed green line shows the computation with no FL at third order, $K^{(3)}_\text{FL}=0$, and the dotted red line
 	shows the power spectrum without frame-lagging at second order, $K^{(2)}_\text{FL}=0$.
 	\label{fig:pkNoFL}}
 	\end{center}
 \end{figure}

The SPT power spectrum is constructed as 
\begin{align}\label{PcbcbSPT}
P_{cb}^\text{SPT}(k) &= P^L_{cb}(k) + P^{22}_{cb}(k) + P^{13}_{cb}(k),  
\end{align}
with 1-loop contributions
\begin{align}
P^{22}_{cb}(k)  &=  2 \int \Dk{p} \big[ F_2(\vk-\vp,\vp) \big]^2  P^L_{cb}(|\vk-\vp|)  P^L_{cb}(p), \label{P22}\\
P^{13}_{cb}(k)  &=  6   P^L_{cb}(k) \int \Dk{p} F^s_3(\vk,-\vp,\vp)  P^L_{cb}(p). \label{P13}
\end{align}

To see the importance of the frame-lagging contributions, and how they bring 
$P_\text{loop}/P_\text{L} \propto k^2$ at large scales, in figure \ref{fig:pkNoFL} we show the ratio of the full SPT loop contributions power spectrum to the linear power spectrum, for 
massive neutrinos with $M_\nu=0.4\,\text{eV}$, including the
frame lagging terms (blue solid line), together with a power law $\propto k^2$ (dashed gray). The green dot-dashed line shows the computation by setting $K^{(3)}_\text{FL}=0$, and the red dotted
line the power spectrum without second order FL,  $K^{(2)}_\text{FL}=0$, but keeping the third order frame-lagging term; the two latter cases have UV divergences which manifest 
in a deviation of a $k^2$ behavior at large scales, while the full power spectrum does tend to $P_\text{loop} \propto  k^2 P_\text{L}$ at very large scales. 
A similar plot is shown in ref.~\cite{Blas:2014hya} (figure 8), to show that the approximation $\delta_{\nu}=\delta^{(1)}_{\nu}$ violates momentum 
conservation. 

We take a closer look to the results of figure \ref{fig:pkNoFL}, when no frame-lagging are considered. By setting $K^{(2)}_\text{FL}=0$, 
eq.~\eqref{D2limitk0} does not scale as $k^2$ but it tends to a constant value, which makes
\begin{equation}\label{PloopKF20}
P^\text{loop}_{cb}(k\rightarrow 0) \big|_{K^{(2)}_\text{FL}=0} \sim \frac{9}{98} \underset{p \gg k}{\int} \frac{dp}{4 \pi^2} p^2 P^2_L(p) \left[ \mA(-\vp,\vp)\big|_{\text{No FL}} - \mB(-\vp,\vp)  \right]^2, 
\end{equation}
a constant --- note that it is the frame-lagging term that makes $\mA = \mB$ for planar collapse. 
This explains the large scales behavior $P^\text{loop}/P^L \propto k^{-n_s}$, with $n_s$ the primordial spectral index,  for the red dotted curve in figure \ref{fig:pkNoFL}. Notice that this contribution
comes entirely from $P_{22}$, more specifically from the term $k_iL_i^{(2)}$ in the $F_2$ function of eq.~\eqref{LtoF2}; on the other hand, with the frame-lagging,  $P_{22}$ scales as $k^4$, as follows from eq.~\eqref{D2limitk0}.

For $K^{(3)}_\text{FL}=0$, the analysis is more challenging because the large-scale behavior becomes dominated by the third order LPT kernel.  We numerically obtain that  
\begin{equation}\label{PloopKF30}
P^\text{loop}_{cb}(k\rightarrow 0) \big|_{K^{(3)}_\text{FL}=0} \propto  P_L(k)  \underset{p \gg k}{\int} \frac{dp}{4 \pi^2} p^2 P_L(p),
\end{equation}
which is expected for eq.~\eqref{D3LS} tending to a constant, instead of behaving as $k^2/p^2$. 
The situation is worse than in the previous case, because here the result formally diverges for linear power spectra, $P_L(p) \propto p^n$ at high $p$, for $n\geq -3$. For the case of eq.~\eqref{PloopKF20}, instead, 
the UV divergence appears for $n\geq -3/2$.  The frame-lagging terms tame these UV divergences, rendering them to $n>1/2$ and $n>-1$, for $P_{22}$ and $P_{13}$ respectively, which is a known result in SPT.
Hence, without frame-lagging terms the theory poses UV divergences due to a failure of short-modes cancellations. Only when these are considered, the theory is  well posed and large and small scales decouple.

One may be worried about the precise cancellations between $P_{22}(k)$ and $P_{13}(k)$ that occur at high-$k$. However, these are provided only by the terms containing linear displacement field kernels 
in eqs.~\eqref{LtoF2} and \eqref{LtoF3}, so they cancel in the same manner as in the $\Lambda$CDM. Technical difficulties, particularly for numerical integration, arise because $P_{22}$ has IR divergences not only when the internal momentum is equal to zero, but also when its magnitude is
equal to the external momentum; see, e.g. \cite{Carrasco:2013sva}. 

\vspace{0.5cm}

Now, coming back to eq.~\eqref{deltanuProxy}, we further take 
\begin{equation}\label{alphaApp}
 \alpha(k) = \frac{\delta^{(1)}_\nu(k)}{\delta^{(1)}_{cb}(k)} \simeq \frac{T_{\nu} (k)}{T_{cb} (k)},   
\end{equation}
where the equality holds true for adiabatic perturbations, being $T_\nu(k,z)$ and $T_{cb}(k,z)$ the transfer functions 
for neutrinos and the $cb$ fluid, that relate the amplitude of linear density fields from their primoridial initial state set by inflation up to redshift $z$. 
Hence, function $A(k)$, given by eq.~\eqref{Aofk}, becomes
\begin{equation} \label{Anu}
 A(k,t) = 4\pi G \bar{\rho}_m \left(f_{cb}  + f_{\nu} \frac{ T_{\nu}(k,t)}{T_{cb}(k,t)} \right).
\end{equation}
To compare our method with others approximations followed in the literature, we compute the growing function $D_+(k,t)$ from eq.~\eqref{DplusEv} 
and use it to evolve a linear power spectrum obtained from the code \texttt{CAMB} \cite{Lewis:1999bs} at $z=10$ up to $z=0$, and compare it with the output of \texttt{CAMB} at $z=0$. 
In figure \ref{fig:pkEvolved} we show the relative difference between these two quantities for different cases: 
a) the dashed red line is obtained by evolving the power spectrum
of massive neutrinos, with $M_\nu=0.4\,\text{eV}$, using $A(k,z) = 4\pi G \bar{\rho}_{m}$; b) blue dot-dashed uses $A(k)=A_0= 4\pi G \bar{\rho}_m f_{cb}$; 
c) solid black uses $A(k,z)$ given by eq.~\eqref{Anu}; and, d) green dotted line evolves the massless neutrino case with $A(k,z) = 4\pi G \bar{\rho}_m$. 
We note that the for quasi-linear scales the approximation given by eq.~\eqref{Anu} is very accurate.
On the other hand, the relative error of about 1\%  at very large scales is due to relativistic contributions to the Poisson equation, suppressed by factors $(aH/k)$.  
Such effect introduces an additional scale dependence which is not accounted for in eq.~\eqref{DplusEv}, but it does in Einstein-Boltzmann codes that compute the linear power spectrum.
Notice that we will not use $D_+(k,t)$ to evolve linear fields, which are obtained directly 
from \texttt{CAMB}; however, both $A(k,t)$ and $D_+(k,t)$ are used to obtain the loop corrections to matter and tracer statistics.  

 \begin{figure}
 	\begin{center}
 	\includegraphics[width=3.4 in]{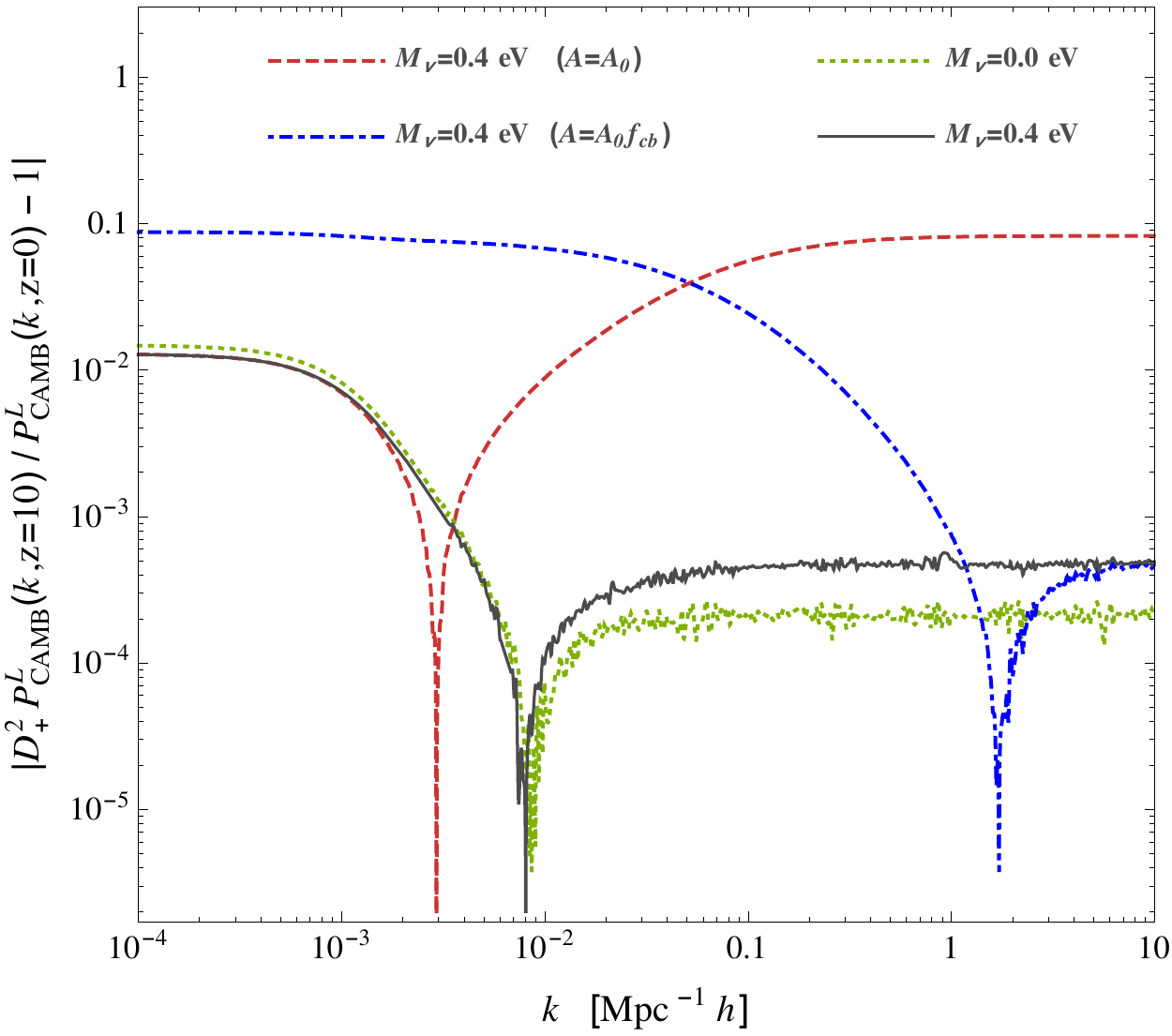}
 	\caption{Relative differences of evolved auto power spectra of the $cb$ fluid from redshift $z=10$ to $z=0$. The evolution is given by solving
 	eq.~\eqref{DplusEv}, such that we compare $\big[D_+(k,z=0)/D_+(k,z=10)\big]^2 P_L(k,z=10)$ with  $P_L(k,z=0)$, where both linear power spectra are obtained from \texttt{CAMB}. 
 	The dashed red line is obtained by evolving the power spectrum
 	of massive neutrinos using $A(k,z) = 4\pi G \bar{\rho}_{m} = A_0$; blue dot-dashed uses $A(k,z)= 4\pi G \bar{\rho}_{cb} = A_0 f_{cb}$; solid black, $A(k,z)$ given by eq.~\eqref{Anu}; and green dotted line evolves the massless neutrino case. 
 	\label{fig:pkEvolved}}
 	\end{center}
 \end{figure}

 \begin{figure}
 	\begin{center}
 	\includegraphics[width=6 in]{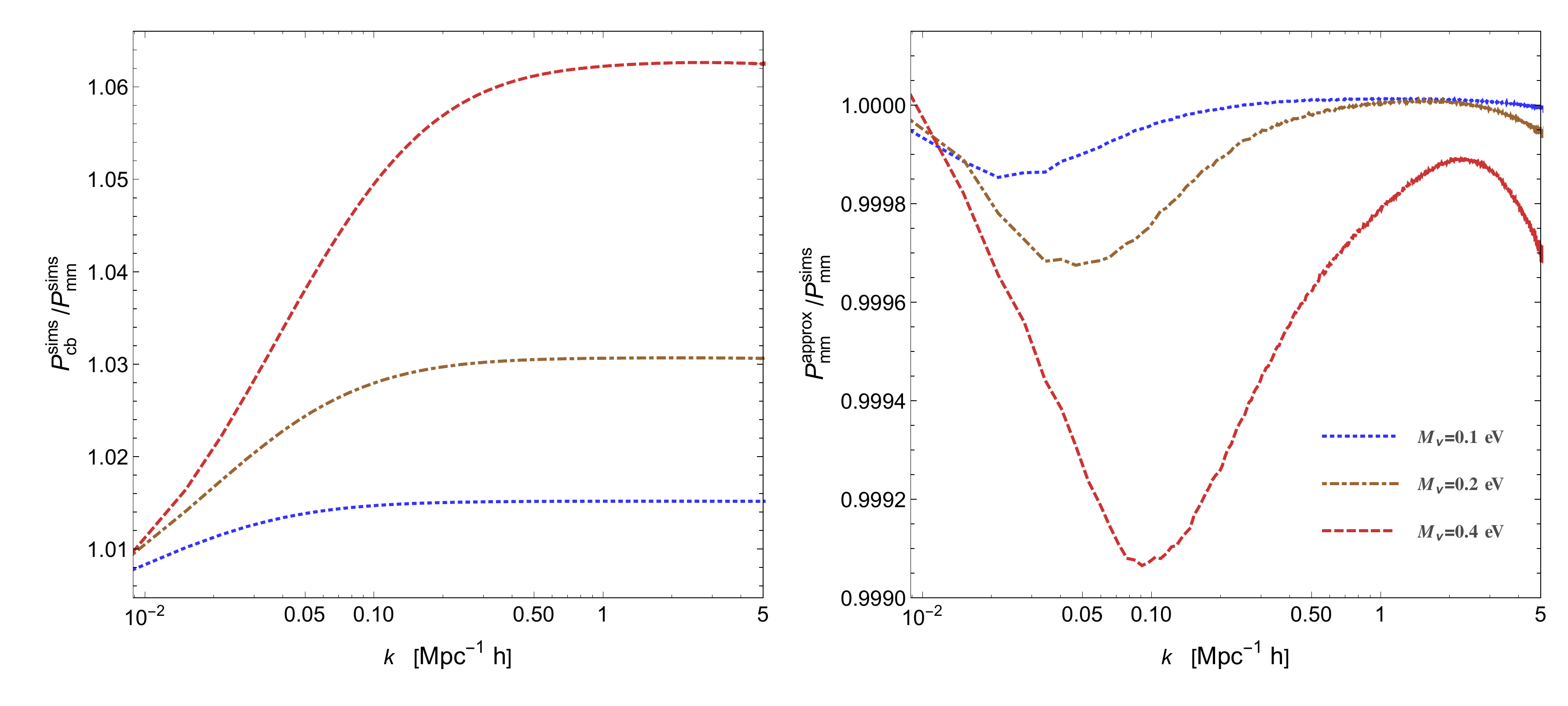}
 	\caption{Comparison of non-linear power spectra for the different cases considered in this work: $M_\nu=0.1,\,0.2,\,0.4$ eV at redshift $z=0.5$; shown in 
 	dotted blue, dot-dashed brown and dashed red lines. 
 	The left panel show the ratios of the $cb$-$cb$ power spectrum to the $m$-$m$ power spectrum, both obtained from the simulations. The right panel shows the ratio of the approximated $m$-$m$ power spectrum, given by eq.~\eqref{Pmmapprox}, to the $m$-$m$ power spectrum  obtained from the simulations. The difference between the latter is smaller than the $0.1 \%$ for all models considered here up to
 	$k=5\, \text{Mpc}^{-1} \,h$, showing that the approximation $\delta_\nu= (T_\nu/T_{cb}) \delta_{cb}$ works also at highly non-linear scales.
 	\label{fig:PmmApprox}}
 	\end{center}
 \end{figure}

We now test how good is the approximation $\delta_\nu= (T_\nu/T_{cb}) \delta_{cb}$, and what is its range of validity.
The $m$-$m$ power spectrum can be decomposed as
\begin{equation} \label{Pmm}
P_{mm}(k) = f_{cb}^2 P_{cb}(k) + 2 f_{cb} f_\nu P_{cb,\nu} + f_\nu^2 P_{\nu}(k),     
\end{equation}
where  $P_{cb,\nu}$ is the cross-power spectrum of $cb$ and neutrino fields, and  $P_{\nu}$ the auto-power spectrum of neutrinos. Under our approximation for function $\alpha(k)$ given by eq.~\eqref{alphaApp}, the latter two are given by $P_{cb,\nu} = (T_\nu/T_{cb}) P_{cb}$ and $P_{\nu} = (T_\nu/T_{cb})^2 P_{cb}$, which can be used to approximate $P_{mm}$ as
\begin{equation}\label{Pmmapprox}
P_{mm}^\text{approx}(k) = \left[ f_{cb}^2 + 2 f_\nu f_{cb} \left(\frac{T_\nu(k)}{T_{cb}(k)}\right)  + f_\nu^2  \left(\frac{T_\nu(k)}{T_{cb}(k)}\right)^2  \right] P_{cb}(k).
\end{equation}

In the left panel of figure \ref{fig:PmmApprox} we plot the ratio of nonlinear power spectrum $cb$ to the nonlinear power spectrum of the total matter field ($cb\, +\, \nu$), both obtained directly from the \textsc{Quijote} suite of simulations (below, in the following section, we briefly describe the specifications of these simulations). We are doing this for cosmologies with massive neutrinos $M_\nu=0.1,\,0.2,\,0.4$ eV corresponding to $f_{cb}=0.9925,\, 0.985,\, 0.97$, and consider redshift $z=0.5$. At very large scales the matter power spectrum, $P^\text{sims}_{mm}$, for the three models tend to $P^\text{sims}_{cb}$ because neutrino and $cb$ overdensities behave equally. On the other hand, at small scales neutrinos do not cluster, and $P^\text{sims}_{mm}$ become suppressed by factors $f_{cb}^2$, tending to $f_{cb}^2 P^\text{sims}_{cb}$, so the ratios go to the constants $1/f_{cb}^2$. 

In the right panel of figure \ref{fig:PmmApprox} we plot the ratios of the approximation given in eq.~\eqref{Pmmapprox}, with $P_{cb}=P^\text{sims}_{cb}$ obtained from the simulations, to the matter power spectrum $P^\text{sims}_{mm}$. These two power spectra differ by less than $0.1\,\%$ over the interval $k\in (0.009,5) \, \text{Mpc}^{-1}\, h$ for all considered models. This analysis shows that the approximation given by eq.~\eqref{alphaApp} is valid well inside the non-linear regime.

\end{section}

\begin{section}{Real space correlation function} \label{sect:RSCF}

In this section, we construct the real space correlation function for tracers within the CLPT framework, using the LPT for CDM in the presence of massive neutrinos developed in the previous sections. 
Here, we will compare our analytical results only to simulated particles, both $cb$ and total matter.  A comparison to CDM halos is performed in section \ref{sect:halos}.

We will assume the existence of a Lagrangian biasing function $F$ that relates the density fluctuations of tracers $\delta_X(\vq)$ with a set of operators constructed out of 
the CDM Lagrangian overdensities. Our biasing scheme is simple since we introduce only local and curvature biases, which shows to provide the level of accuracy necessary to match the simulations we consider.
If desired, tidal bias can be introduced along the lines of ref.~\cite{Vlah:2016bcl}, with small modifications due to the generalized kernels used \cite{Valogiannis:2019nfz}. Hence, $cb$ and tracer initial densities are related by 
\begin{equation}\label{LagrangianF}
 1+\delta_X(\vq) = F(\delta_{cb}, \nabla^2 \delta_{cb}) = \int \frac{d^2\mathbf{\Lambda}}{(2\pi)^2} \tilde{F}(\mathbf{\Lambda}) e^{i \mathbf{D}\cdot \mathbf{\Lambda}}. 
\end{equation}
In the second equality $\tilde{F}(\mathbf{\Lambda})$ is the Fourier transform of $F(\ve D)$, with arguments $\mathbf{D}=(\delta_{cb}, \nabla^2 \delta_{cb})$ and spectral parameters $\mathbf{\Lambda}=(\lambda,\eta)$, dual to $\mathbf{D}$.
Assuming number conservation of tracers, $\big[1+\delta_X(\vx) \big]d^3x = \big[1+\delta_X(\vq) \big]d^3q$, one obtains
\begin{equation}\label{PsitodeltaX}
 1+\delta_X(\vx) = \int \Dk{k} \int d^3q e^{i \vk \cdot (\vx -\vq)} \int \tilde{F}(\mathbf{\Lambda}) e^{i \mathbf{D}\cdot \mathbf{\Lambda} - i \vk \cdot \mathbf{\Psi}},
\end{equation}
which evolves initially biased tracer densities using the map of eq.~\eqref{qTox} between Lagrangian and Eulerian coordinates.    
Renormalized bias parameters are obtained through \cite{Matsubara:2008wx,Aviles:2018thp}
\begin{equation}\label{renormBias}
 b_{nm} = \int \frac{d\mathbf{\Lambda}}{(2\pi)^2} \tilde{F}(\mathbf{\Lambda}) e^{-\frac{1}{2} \mathbf{\Lambda}^\text{T} \mathbf{\Sigma} \mathbf{\Lambda} } (i \lambda)^n (i \eta)^m,
\end{equation}
with covariance matrix components $\Sigma_{11} = \langle \delta_{cb}^2 \rangle$, $\Sigma_{12} =\Sigma_{21} = \langle \delta_{cb} \nabla^2 \delta_{cb} \rangle$ and $\Sigma_{22} = \langle (\nabla^2\delta_{cb})^2 \rangle$.
We identify $b_n =b_{n0}$ with the local bias parameter of order $n$, and $b_{\nabla^2\delta} = b_{01}$ with the curvature bias parameter. 
The correlation function $\xi_{X}(r)$ for tracer $X$ is obtained from eqs.~\eqref{LagrangianF}, \eqref{PsitodeltaX} and \eqref{renormBias} by using 
the standard methods of CLPT \cite{Carlson:2012bu,Vlah:2015sea,Uhlemann:2015hqa,Vlah:2018ygt,Aviles:2018thp},
\begin{align}\label{CLPTxicb}
&1+\xi_{X, cb}(r) = \int  \frac{d^3 q}{(2 \pi)^{3/2} |\mathbf{A}_L|^{1/2}} e^{- \frac{1}{2}(\ve r-\vq)^\mathbf{T}\mathbf{A}_L^{-1}(\ve r-\vq) } 
 \Bigg\{ 1 - \frac{1}{2} A_{ij}^{loop}G_{ij} +\frac{1}{6}\Gamma_{ijk}W_{ijk} \nonumber\\
&\quad + b_1 (-2 U_i g_i - A^{10}_{ij}G_{ij}) + b_1^2 (\xi_L - U_iU_jG_{ij}- U_i^{11}g_i) + b_2(\frac{1}{2} \xi_L^2 -U_i^{20}g_i - U_iU_jG_{ij}) \nonumber\\
&\quad - 2 b_1 b_2 \xi_L U_i g_i  + 2(1+b_1) b_{\nabla^2 \delta} \nabla^2 \xi_L 
       + b^2_{\nabla^2 \delta} \nabla^4 \xi_L \Bigg\}, 
\end{align}
where we are using the label ``$cb$'' in $\xi_{X, cb}$ to distinguish that we are biasing the $cb$ fluid and not the whole matter density $\delta_{m}=f_{cb} \delta_{cb}+ f_\nu \delta_{\nu}$.
The matrix 
$A^L_{ij}(\vq) = \langle \Delta_i^{(1)} \Delta_j^{(1)} \rangle_c $, with $\Delta_i = \Psi_i(\vq_2) - \Psi_i(\vq_1)$, is the correlation of the difference of linear displacement fields for initial positions 
separated by a distance $\vq=\vq_2-\vq_1$,
\begin{equation}\label{ALij}
 A^L_{ij}(\vq) = 2 \int \Dk{p} \big( 1 - e^{i\vp\cdot \vq} \big)\frac{p_i p_j}{p^4} P^L_{cb}(p), 
\end{equation}
and  the tensors $g_i = (\mathbf{A}_L^{-1})_{ij}(r_j - q_j)$, $G_{ij} = (\mathbf{A}_L^{-1})_{ij} - g_ig_j$, and $\Gamma_{ijk} = (\mathbf{A}_L^{-1})_{\{ij} g_{k\}} - g_ig_jg_k$. 
We further use the linear correlation function 
\begin{equation}
\xi_L(q) = \int \Dk{p} e^{i \vp \cdot \vq}   P^L_{cb}(p),
\end{equation}
and the functions
\begin{align}
W_{ijk} =  \langle \Delta_{i} \Delta_i \Delta_{k}\rangle_c, \qquad
A_{ij}^{mn} = \langle \delta_{cb}^{m}(\vq)\delta_{cb}^{n}(0)\Delta_i \Delta_i \rangle_c, \qquad
U^{mn}_i = \langle \delta_{cb}^{m}(\vq)\delta_{cb}^{n}(0)\Delta_i \rangle_c,
\end{align}
such that $A^\text{loop}_{ij} \equiv A_{ij}^{00}- A^L_{ij}$, and $U_i \equiv U^{00}_i$. For example, the linear piece of function $U_i$ is
\begin{equation}\label{Ui}
 U^L_i(\vq) = - i \int \Dk{p} e^{i\vp\cdot\vq} \frac{p^i}{p^2} P^L_{cb}(p).
\end{equation}

The CLPT correlation function given by eq.~\eqref{CLPTxicb} has the same structure that in the massless neutrino $\Lambda$CDM model. 
The differences with the $f_{\nu}=0$ case appear through the functions  $U_i, \,A_{ij},\, W_{ijk}$, since they are ultimately constructed out of the LPT kernels. 
In Appendix \ref{app:kq} we show how these reduce to integrals of the kernels and linear power spectra.

The correlation function for tracers can be obtained as well by considering the bias of the auto-correlation function with respect to the total matter field, $\xi_{X, m}(r)$, that we approximate by replacing 
$P^L_{cb}$ by $P^L_{m} = f_{cb}^2 P^L_{cb} + 2 f_{cb} f_\nu P^L_{cb,\nu} + f_\nu^2 P^L_\nu$ in the linear functions appearing in eq.~\eqref{CLPTxicb} ($A^L_{ij}$, $U^L_i$
and $\xi_L$) and multiplying by $f_{cb}^2$ all loop contributions.\footnote{An alternative is to use 
$\xi_{X,m}(r) = f_{cb}^2 \xi^\text{CLPT}_{X,cb}(r) + (1+b_1)^2 \big( 2 f_{cb}f_{\nu} \xi^\text{ZA}_{cb,\nu}(r)  + f_{\nu}^2 \xi^\text{ZA}_{\nu}(r) \big)$,
where the Zeldovich approximation-like correlation functions $\xi^\text{ZA}_{cb,\nu}(r)$ and $\xi^\text{ZA}_{\nu}(r)$
are obtained by taking only the ``1'' term inside the brackets of eq.~\eqref{CLPTxicb} and substituting $P_{cb}$ by
$P_{cb,\nu}$ and $P_\nu$, respectively, in eq.~\eqref{ALij}. Both approaches yield similar results, with differences smaller than the $1\%$ at all scales.} 
Our approach is analogous to the usually followed for the SPT power spectrum, that approximates $P_{m}^\text{SPT}= f_{cb}^{2} P_{cb}^\text{SPT} + 2 f_{cb}f_\nu P_{cb,\nu}^L + f_\nu^2 P_{\nu}^L$.
By doing so, we neglect the loop contributions in the correlation function coming from non-linear terms of $\delta_\nu$. Although our method is in apparent inconsistency with the general treatment given in the previous sections, 
these loops contributions are smaller than those coming from CDM densities and further suppressed by factors $f_\nu$, hence the error we are committing is very small
as long as the neutrino masses are not very large.

 \begin{figure}
 	\begin{center}
 	\includegraphics[width=6 in]{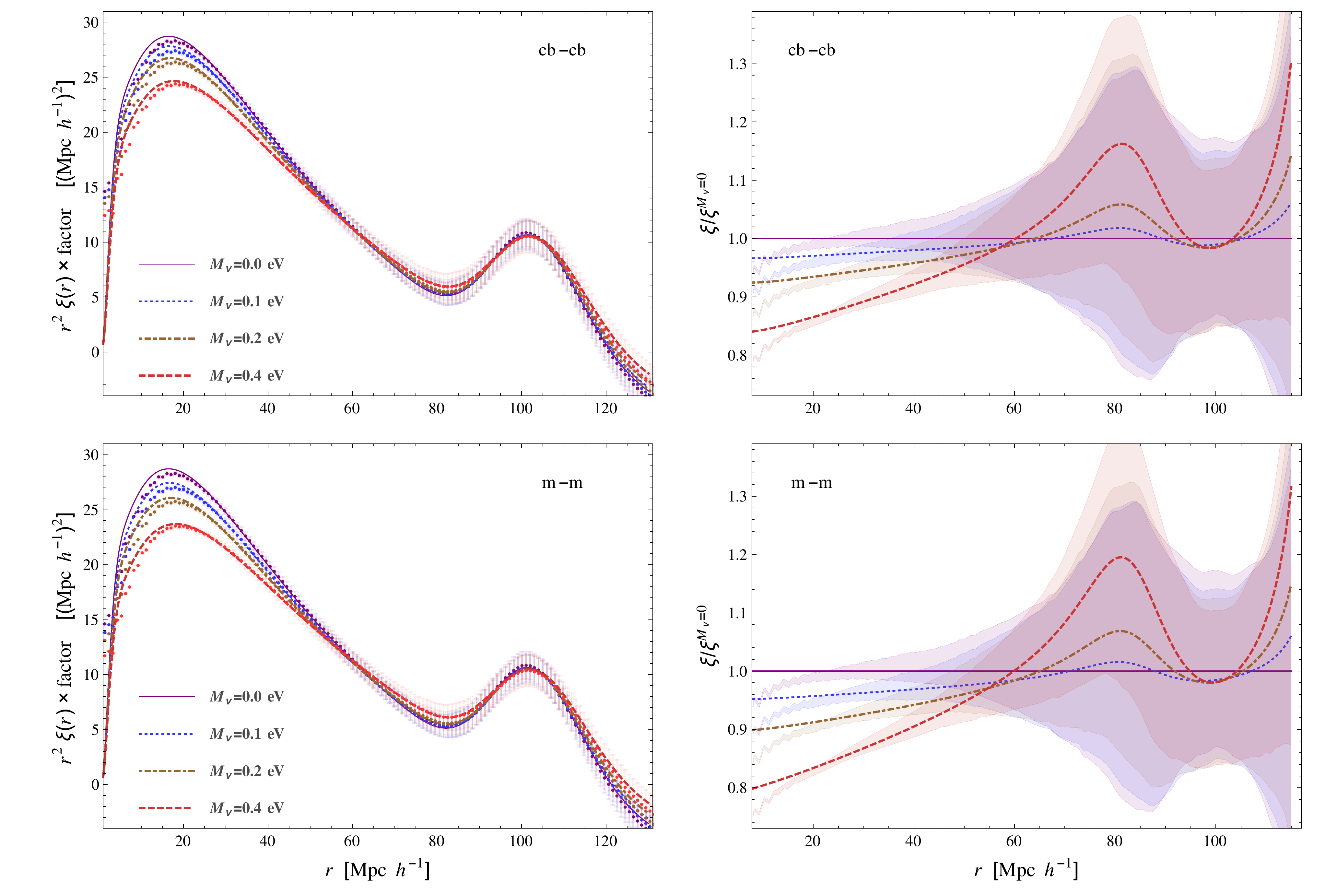}
 	\caption{Particles real space correlation function at $z=0.5$ for $M_\nu=0.0\,\text{eV}$ (solid purple line), $M_\nu=0.1\,\text{eV}$ (dotted blue), $M_\nu=0.2\,\text{eV}$ (dot-dashed brown) and 
 	 $M_\nu=0.4\,\text{eV}$ (dashed red) compared to $N$-body simulations data (dots). The upper figures show the $cb$ fluid auto-correlation functions and the lower figures the $m= cb +\nu$ auto-correlation functions.
 	For visualization purposes, we have multiplied the results by constant factors 
 	$P_L^{M_\nu=0.4}(k_0)/P_L^{M_\nu=0,0.1,0.2,0.4}(k_0)$, with $k_0=10^{-4}$, such that at large scales the 
 	corresponding power spectrum in all models have approximately the same amplitude. The right panels show the ratios over the massless neutrino case, with the shaded regions the 
 	simulated data RMS errors.
 	\label{fig:CFmatter}}
 	\end{center}
 \end{figure} 

We want to assess the goodness of our analytical model by comparing directly to the particles of $N$-body simulations. Later, in section \ref{sect:halos}, we will compare to tracers. 
To this end, we use measurements from the \textsc{Quijote} $N$-body simulation suite \cite{Villaescusa-Navarro:2019bje}. The fiducial cosmology in the \textsc{Quijote} suite has
$\Omega_m =0.3175$, $\Omega_b=0.049$, $h=0.6711$, $n_s=0.9624$, and $\sigma_8=0.834$, and $M_\nu=0$ eV. There are also three massive neutrino cosmologies (assuming three degenerate massive neutrinos) with total mass $M_\nu=0.1,0.2,0.4$ eV 
corresponding to $f_\nu=0.0075,\, 0.015,\, 0.03$, respectively. The simulations volume is $(1\hgpc)^3$, and uses $512^3$ CDM particles for the massless neutrino cosmology, and $512^3$ CDM 
and $512^3$ neutrino particles for the massive neutrino cosmologies. Note that in these simulations $\sigma_8$ is kept fixed, such that the primordial amplitude $A_s$ is different for each model. 
To reduce the sample variance in the simulation measurements, we use $100$ realizations at each cosmology.

By comparing to the simulated particles we get a direct test of our theory
since in this case we have no free parameters. To this end we set all bias parameters to
zero and perform the integral in eq.~\eqref{CLPTxicb}. We show the analytical results together with the simulated data in figure \ref{fig:CFmatter} for the CDM particles ($cb$-$cb$) in the top panels, 
and to all particles ($m$-$m$), including also the massive neutrinos, in the bottom panels.  
The differences among the models are dominated by their large scale, primordial amplitudes; hence, to isolate the effects of late time clustering, we have multiplied the particle real space correlation functions by constants equal to 
$\text{``factor''}= P_L^{M_\nu=0.4}(k_0)/P_L^{M_\nu=0,0.1,0.2,0.4}(k_0) = 1.29,1.22,1.14,1$, with $k_0=10^{-4} \, \text{Mpc}^{-1} h$, 
such that the corresponding power spectrum in all models have approximately the same primordial amplitude $A_s=2.74 \times 10^{-9}$. 
The figures on the right column show the ratios $\xi/\xi^{M_\nu=0}$ 
(including the constant factors)
of the different massive cases to the massless neutrino correlation function, with the shaded region showing the RMS error of the simulated data. We find that our analytical approach 
show the same level of accuracy for all models, being consistent with the data down to $r=20\, \text{Mpc} \, h^{-1}$, being this the standard level of precision provided by CLPT \cite{Carlson:2012bu,Vlah:2015sea}.
Below this scale the predictions of CLPT overshoot the $N$-body simulated data.


  \end{section}

\begin{section}{Redshift-space correlation function}\label{sect:zSCF} 

In this section we turn our attention to the effects of RSD in the 2-point statistics. As before, 
we will present the whole theory for biased tracers and we will compare the analytical results only to simulated particles. 
In section \ref{sect:halos}, we will compare to CDM halos.

An object located at a comoving real space position $\vx$
is observed to be at an apparent, redshift-space position $\ve s$, due to the Doppler effect induced by its
peculiar velocity, $a \mathbf{\dot{\Psi}}$, relative to the Hubble flow. Hence, both coordinate systems
are related by $ \ve s = \vx + \ve u$, with ``velocity'' $\ve u$ defined as 
\begin{equation}\label{velu}
\ve u \equiv  \vhn \frac{\mathbf{\dot{\Psi}} \cdot \vhn}{H},
\end{equation}
where we adopted the plane-parallel approximation, for which $\vhn$ is a constant vector in the direction of the survey, instead of being equal to the 
position unit vector $\hat{\vx}$. 
The map between Lagrangian coordinates and redshift space Eulerian positions becomes 
\begin{equation}\label{RSDmap}
 \ve s = \vq + \mathbf{\Psi} + \vhn \frac{\mathbf{\dot{\Psi}} \cdot \vhn}{H}.
\end{equation}
Conservation of number of objects, $\big[1+\delta_s(\ve s)\big]d^3s = \big[1+\delta(\vx)\big]d^3x$, yields
\begin{equation}
 (2\pi)^3\delta_\text{D}(\vk) + \delta_s(\vk) = \int d^3x \big[1+\delta(\vx)\big] e^{i \vk \cdot(\vx+ \ve u(\vx))},  
\end{equation}
and the redshift-space correlation function becomes \cite{Scoccimarro:2004tg}
\begin{equation}\label{RSDCF}
  1 + \xi_s(\ve s) = \int \Dk{k} d^3x \,e^{i\vk\cdot(\ve s - \vx)} \Big[ 1+\mathcal{M}(\vk,\vx) \Big], 
\end{equation}
with pairwise velocity generating function  
\begin{equation}\label{VDgenF}
1+\mathcal{M}(\vk,\vx) =  \left\langle \big(1+\delta_1 \big)\big(1+\delta_2\big)  e^{i \vk \cdot \Delta \ve u}  \right\rangle,
\end{equation}
where $\Delta \ve u = \ve u(\vx_2)-\ve u(\vx_1)$, $\ve x = \vx_2 - \vx_1$ and $\delta_1 = \delta_{cb}(\vx_{1})$, $\delta_2 = \delta_{cb}(\vx_{2})$. 
We  expand the pairwise velocity generating function in cumulants as \cite{Scoccimarro:2004tg,Vlah:2018ygt}
\begin{equation} \label{Mcumexp}
 1+\mathcal{M}(\vk,\vx) = \big[ 1 + \xi(x) \big] \exp \left[  i k_i v_{12,i}(\vx) - \frac{1}{2} k_i k_j \sigma^2_{12,ij}(\vx)  + \cdots \right],
\end{equation}
with $ \xi(x)$ the real space correlation function, $\mathbf{v}_{12}$ the pairwise velocity  and $\mathbf{\sigma}^2_{12}$ the pairwise velocity dispersion, with components 
\begin{align}
 v_{12,i}(\vx)&=  \frac{\langle (1+\delta_1)(1+\delta_2) \Delta u_i \rangle_c }{1 + \xi(x)}, \\
 \sigma^2_{12,ij}(\vx) &=  \frac{\langle (1+\delta_1)(1+\delta_2) \Delta u_i \Delta u_j \rangle_c }{1 + \xi(x)} - v_{12,i}(\vx)v_{12,j}(\vx).
\end{align}
To be consistent in including all 1-loop contributions, one should also consider the third and fourth cumulant of the pairwise velocity generating function. However, by keeping only up to the second 
cumulant, as in eq.~\eqref{Mcumexp}, the $\vk$-integral in eq.~\eqref{RSDCF} can be performed analytically. By doing so, one obtains \cite{Vlah:2018ygt}
\begin{equation} \label{GSMcf}
 1 + \xi_s(\ve s) = \int \frac{d^3 x}{(2\pi)^{3/2} |\mathbf{\sigma}^2_{12}|^{1/2}}
 \big[1+\xi(x)\big] \exp \left[ -\frac{1}{2} (\ve s -\vx -\mathbf{v}_{12} ) [\mathbf{\sigma}^2_{12}]^{-1}(\ve s -\vx -\mathbf{v}_{12} )  \right], 
\end{equation}
which is the GSM expression for the redshift-space correlation function \cite{Fisher:1994ks,Scoccimarro:2004tg,Reid:2011ar}.

  \begin{figure}
 	\begin{center}
 	\includegraphics[width=6 in]{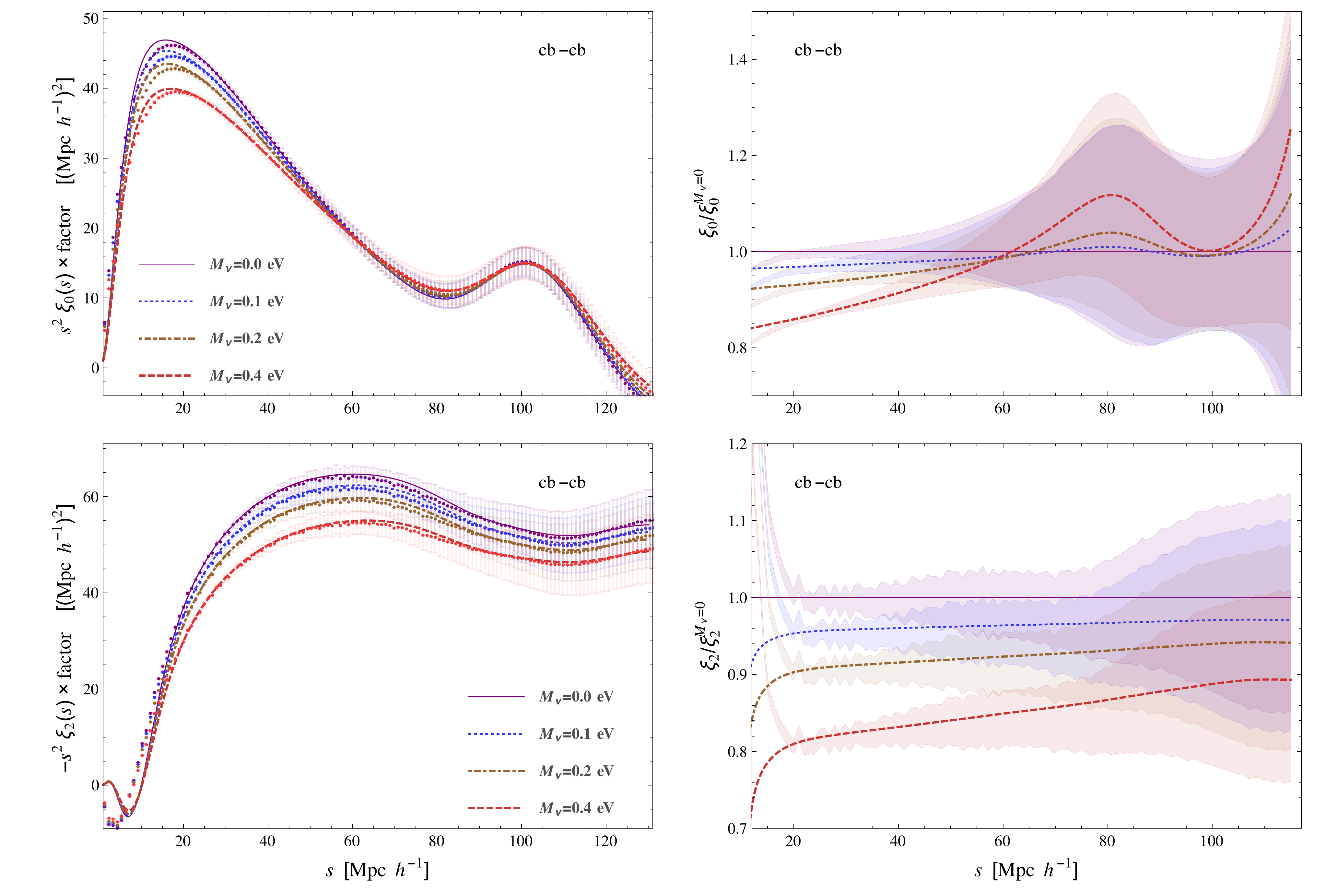}
 	\caption{$cb$ particles redshift space correlation functions at $z=0.5$. The upper figures show the monopole and the lower figures the quadrupole.
 	The right panels show the ratios over the massless case, with the shaded regions the 
 	simulated data RMS errors. We use an EFT parameter $\alpha_\sigma=13.5/f_0^2 \times (\hmpc)^2$ in all cases.
 	\label{fig:CFmultipolescb}}
 	\end{center}
 \end{figure}

The method to obtain expressions for the velocity and velocity dispersion is very similar as in the $\Lambda$CDM case. However one should consider that in the presence of massive neutrinos, the growth function $D_+$ is scale dependent,
and hence the logarithmic growth factor 
\begin{equation}
f(k,t) = \frac{d \ln D_+(k,t)}{d \ln a(t)}, 
\end{equation}
also becomes scale dependent. For notational convenience, we define $f_0 \equiv f(k_0,t)$,
with $k_0$ an arbitrary scale that we choose to correspond to a sufficiently long mode, such that $f_0 = f^{M_\nu=0}$. With this, the time derivative of the Lagrangian displacement at perturbative order $n$ can be written as 
\begin{equation}
\dot{\Psi}_i^{(n)}(\vq) = n f_0 H  \underset{\vk_{1\cdots n}= \vk}{\int} \frac{i}{n!}  L^{f\,(n)}_i(\vk_1,\dots,\vk_n;t)  \delta(\vk_1) \cdots \delta(\vk_n),
\end{equation}
with kernels 
\begin{equation}
L^{f\,(n)}_i(\vk_1,\dots,\vk_n) = \frac{f(k_1) + \cdots + f(k_n)}{n f_0} L_i(\vk_1,\dots,\vk_n) + \frac{1}{n f_0 H} \dot{L}^{(n)}_i(\vk_1,\dots,\vk_n). 
\end{equation}
If $f$ is scale independent, and we further use the static kernels approximation, we obtain the standard result $\dot{\Psi}^{(n)} = n f H  \Psi^{(n)}$, widely used for $\Lambda$CDM and exact for EdS kernels.
We employ CLPT to obtain the pairwise velocity and velocity dispersion for the $cb$ fluid, see \cite{Wang:2013hwa,Vlah:2016bcl},
\begin{align}\label{v12}
&\big[1+\xi_{X,cb}(r)\big] v_{12,i}(\ve r)= f_0  \int  \frac{d^3 q\, e^{- \frac{1}{2}(\ve r-\vq)^\mathbf{T}\mathbf{A}_L^{-1}(\ve r-\vq) } }{(2 \pi)^{3/2} |\mathbf{A}_L|^{1/2}} 
 \Bigg\{ -g_r \dot{A}_{ri} - \frac{1}{2} G_{rs} \dot{W}_{rsi} \nonumber\\
&\quad + b_1 \left( 2 \dot{U}_i - 2 g_r \dot{A}^{10}_{ri} - 2 G_{rs} U_r \dot{A}_{si} \right) + b_1^2 \left( \dot{U}^{11}_i  - 2 g_r U_r \dot{U}_i  - g_r \dot{A}_{ri} \xi_L  \right)  \nonumber\\
&\quad + b_2 \left( \dot{U}^{20}_i - 2 g_r U_r \dot{U}_i  \right) + 2 b_1 b_2 \xi_L \dot{U}_i  + 2 b_{\nabla^2\delta} \nabla_i \xi_L   \,
 \Bigg\},
\end{align}
and
\begin{align}\label{s212}
&\big[1+\xi_{X,cb}(r)\big] \sigma^2_{12,ij}(\ve r)= f_0^2  \int  \frac{d^3 q \, e^{- \frac{1}{2}(\ve r-\vq)^\mathbf{T}\mathbf{A}_L^{-1}(\ve r-\vq) } }{(2 \pi)^{3/2} |\mathbf{A}_L|^{1/2}}  
 \Bigg\{  \ddot{A}_{ij} - g_r \ddot{W}_{rij} - G_{rs}\dot{A}_{ri}\dot{A}_{sj} \nonumber\\ 
&\quad  + \alpha_\sigma \delta_{ij}  + 2 b_1 \left( \ddot{A}^{10}_{ij} -  g_r \dot{A}_{r\{i} \dot{U}_{j\}} - g_r U_r \ddot{A}_{ij} \right) 
+ b_1^2 \left( \xi_L \ddot{A}_{ij} + 2 \dot{U}_i \dot{U}_j \right)  + 2 b_2 \dot{U}_i \dot{U}_j  \,
\Bigg\},
\end{align}
with $\xi_{X,cb}(r)$ the CLPT tracers correlation function in eq.~\eqref{CLPTxicb}, and
\begin{align}
&\dot{A}_{ij}^{mn}(\vq) = \frac{1}{f_0 H}\langle \delta^m_1 \delta^n_2 \Delta_i \dot{\Delta}_j \rangle, 
                         \qquad  \ddot{A}_{ij}^{mn}(\vq) = \frac{1}{f_0^2 H^2}\langle \delta^m_1 \delta^n_2 \dot{\Delta}_i \dot{\Delta}_j \rangle, \nonumber\\
& \dot{W}_{ijk} = \frac{1}{f_0 H}\langle  \Delta_i  \Delta_j \dot{\Delta}_k \rangle,  \qquad \ddot{W}_{ijk} = \frac{1}{f_0^2 H^2}\langle  \Delta_i  \dot{\Delta}_j \dot{\Delta}_k \rangle,    \nonumber\\
& \dot{U}^{mn}(\vq) = \frac{1}{f_0 H}\langle \delta^m_1\delta^n_2 \dot{\Delta}_i \rangle, 
\end{align}
and, as before, we omitted to write the superscripts $m,n$ when these are zero; e.g, $\dot{A}_{ij} \equiv \dot{A}^{00}_{ij}$.  
The scale dependence of $f(k)$ is included in the above ``dotted'' functions, and we have factorized the factors $f_0$ to keep the same notation, standard  in the literature, as for the massless neutrinos case. 
In Appendix \ref{app:kq} we show how these ``dotted'' $A$, $U$ and $W$ functions are computed numerically.  

   \begin{figure}
 	\begin{center}
 	\includegraphics[width=6 in]{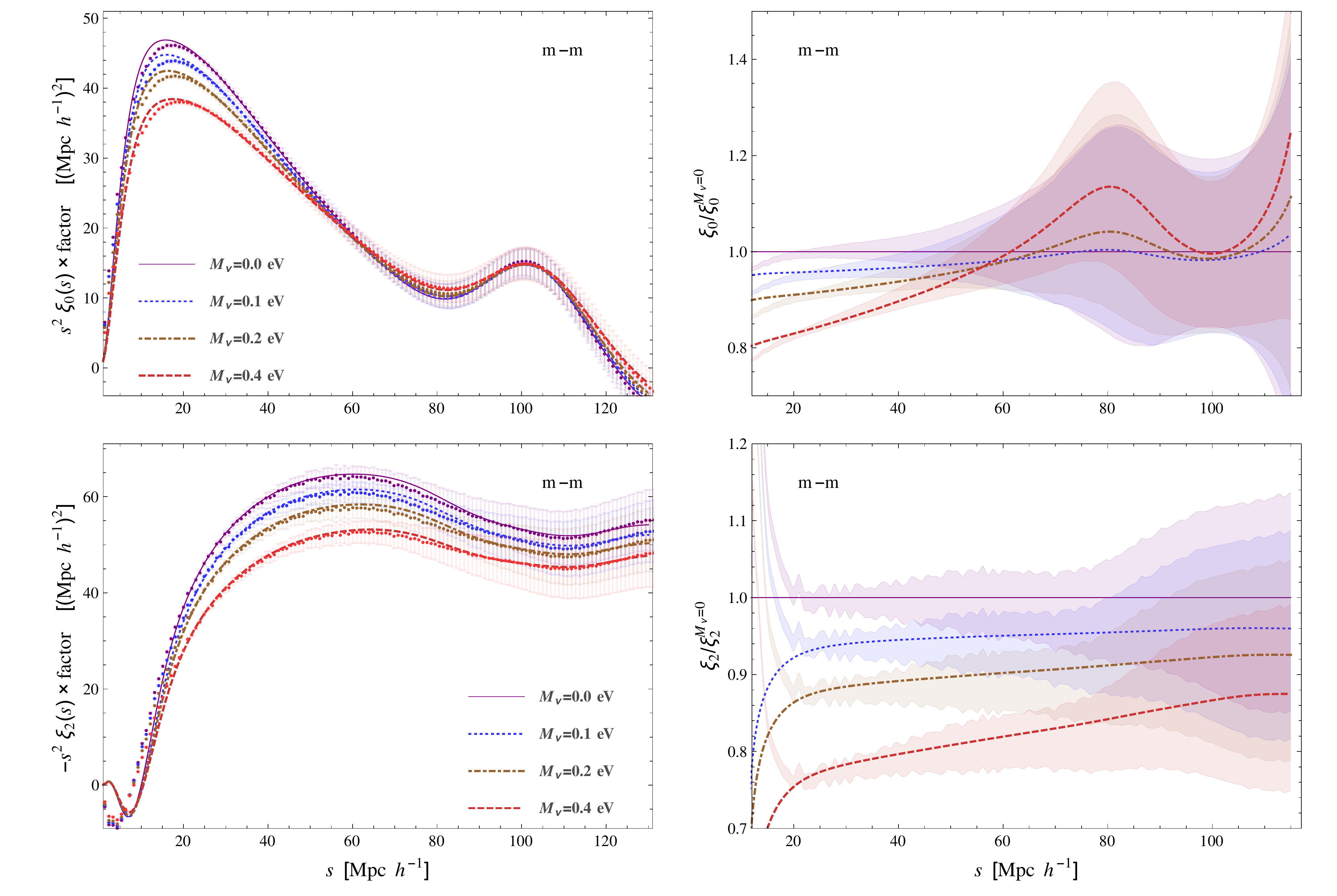}
 	\caption{Total matter ($m=cb+\nu$ particles) redshift space correlation functions. The upper figures show the monopole and the lower figures the quadrupole.
 	The right panels show the ratios over the massless case, with the shaded regions the 
 	simulated data errors. We use EFT parameters  $\alpha_\sigma \times f_0^2/(\text{Mpc}\,h^{-1})^2 =13.5,\,15.5,\,16,\, 17.5$ for $M_\nu=0.0, \, 0.1, \,0.2,\, 0.4 \, \text{eV}$, respectively.
 	\label{fig:CFmultipolesmm}}
 	\end{center}
 \end{figure}

Following \cite{Vlah:2016bcl}, we have included an Effective Field Theory (EFT) counterterm $\alpha_\sigma \delta_{ij}$
to $\ddot{A}_{ij} + 2 b_1 \ddot{A}^{10}_{ij}$, since this combination of functions approach to a non-vanishing, bias-dependent constant at large separation $q$ (times the Kronecker $\delta_{ij}$), 
that is very sensitive to small scale physics, mainly to the zero-lag 
correlator $\langle \dot{\Psi}_i(0)\dot{\Psi}_j(0) \rangle$ which cannot be treated perturbatively. The EFT parameter $\alpha_\sigma$ 
contributes to the pairwise velocity dispersion tensor as
\begin{equation}\label{alphasigma}
 \alpha_\sigma f_0^2 \frac{1+\xi^\text{ZA}_{cb}(r)}{1+\xi^\text{CLPT}_{X,cb}(r)} \delta_{ij}   \in \sigma^2_{12,ij}(\ve r), 
\end{equation}
hence it accommodates well on early works that noticed the necessity of adding a constant shift to match the large scales pairwise velocity 
dispersion observed in $N$-body simulations \cite{Reid:2011ar,Wang:2013hwa}. There are several others EFT counterterms 
entering the CLPT correlation function and the pairwise velocity and velocity dispersion, but they
are either degenerated with curvature bias or subdominant with respect to the contribution of eq.~\eqref{alphasigma} (see the discussion in \cite{Vlah:2016bcl}), 
so in this work we keep only $\alpha_\sigma$. Since this EFT parameter modifies the second cumulant of the pairwise velocity generation function, 
its effect on the redshift space monopole correlation function is small, while the quadrupole is quite sensitive to it, particularly at intermediate scales $r<40\, \text{Mpc} \, h^{-1}$.

 The $cb$ auto correlation function is obtained by substituting eqs.~\eqref{CLPTxicb}, \eqref{v12} and \eqref{s212} into eq.~\eqref{GSMcf}. In figure \ref{fig:CFmultipolescb} we show the monopole (top panels) and quadrupole 
 (bottom panels) of the correlation 
 function for the unbiased case, though we keep the EFT parameter since it is necessary to match the quadrupole simulated data. We have multiplied each correlation function by the same factors as in figure \ref{fig:CFmatter}. The 
 right panels show the ratios to the massless neutrino case with the shaded regions the RMS errors.
 We have used an EFT parameter $\alpha_\sigma = 13.5/f_0^2 \times \text{Mpc}^2\,h^{-2}$, with $f_0=0.76$ for all models. The level of accuracy is similar 
 to that of the correlation function found in figure \ref{fig:CFmatter}, matching the data all the way down to $r = 20 \, \text{Mpc}\,h^{-1}$.
 
 To compare to the total matter simulated data we proceed in an analogous way as we did for the real space correlation function. We substitute $P^L_{cb}$ by $P^L_{m}$ in the leading order functions entering 
 eqs.~\eqref{v12} and \eqref{s212}, and multiply by $f_{cb}^2$ the loop contributions. The comparisons among the theory and simulations are shown in figure \ref{fig:CFmultipolesmm}. The top panels show the 
 monopole of the redshift space correlation function and the bottom panels their quadrupole.
 We have used EFT parameters, reported in $\text{Mpc}^2\,h^{-2}$ units, 
 $\alpha_\sigma=13.5/f_0^2 $ for the massless neutrinos, $\alpha_\sigma=15.5/f_0^2 $ for $M_\nu=0.1\,\text{eV}$, $\alpha_\sigma=16/f_0^2 $ for $M_\nu=0.2\,\text{eV}$, 
 and  $\alpha_\sigma=17.5/f_0^2 $ for $M_\nu=0.4\,\text{eV}$.

\end{section}

\begin{section}{Results for halos}\label{sect:halos}

 \begin{figure}
 	\begin{center}
 	\includegraphics[width=6 in]{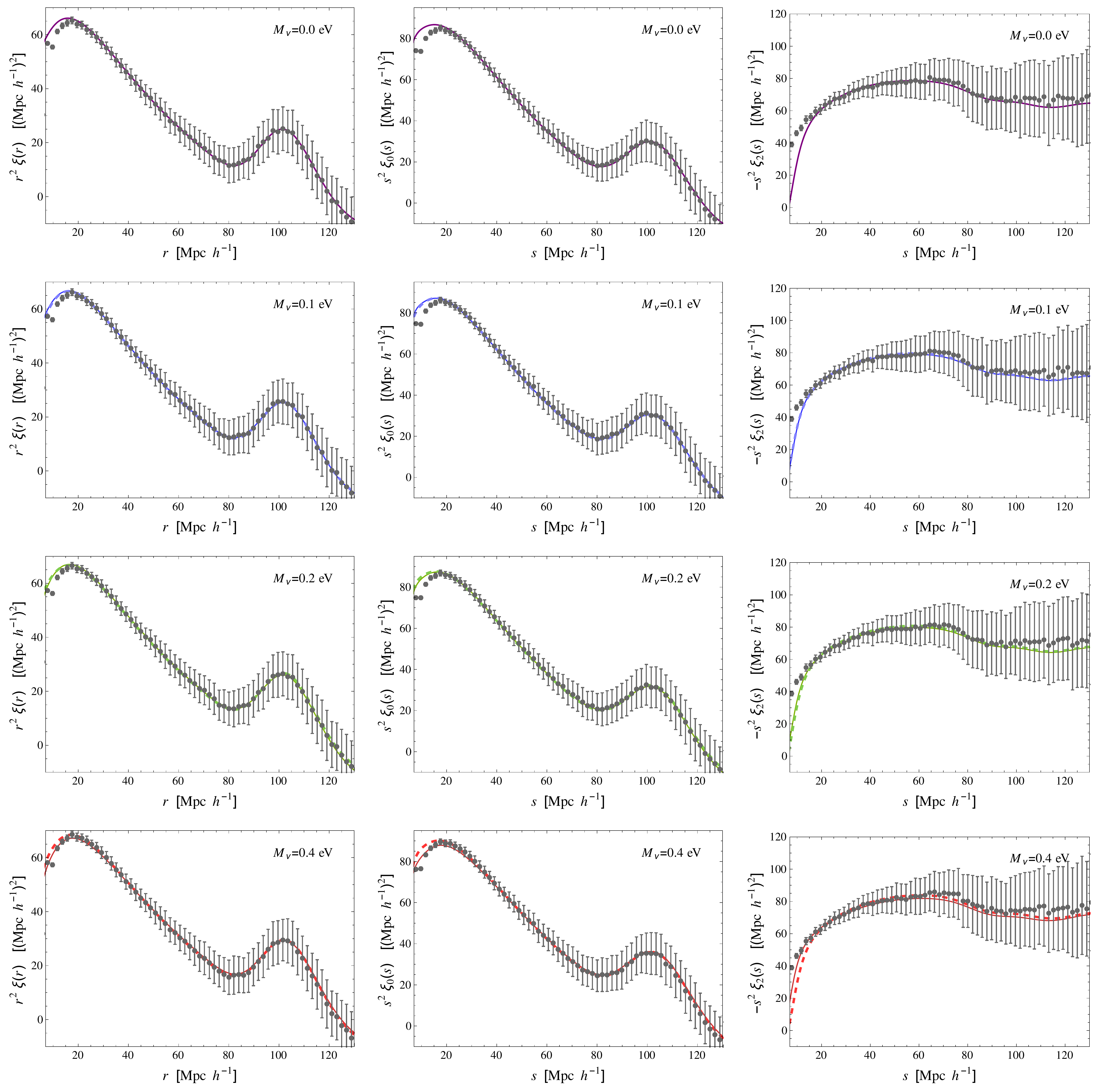}
 	\caption{Halo correlation functions in real space (left column) and $\ell=0,2$ multipoles in redshift space (middle and right columns, respectively). From top to bottom we show the cases $M_\nu=0,\,0.1,\,0.2,\,0.4$ eV. 
 	The solid lines are obtained by applying the
 	biasing scheme to the $cb$ fluid and the dashed lines to the total matter $m= cb +\nu$. 
 	The bias parameters are given in table \ref{tabla:BiasParams}. The relative errors with the simulated data are shown in figure \ref{fig:CFhalosRatios}.
 	\label{fig:CFhalos}}
 	\end{center}
 \end{figure}

 \begin{figure}
 	\begin{center}
 	\includegraphics[width=6 in]{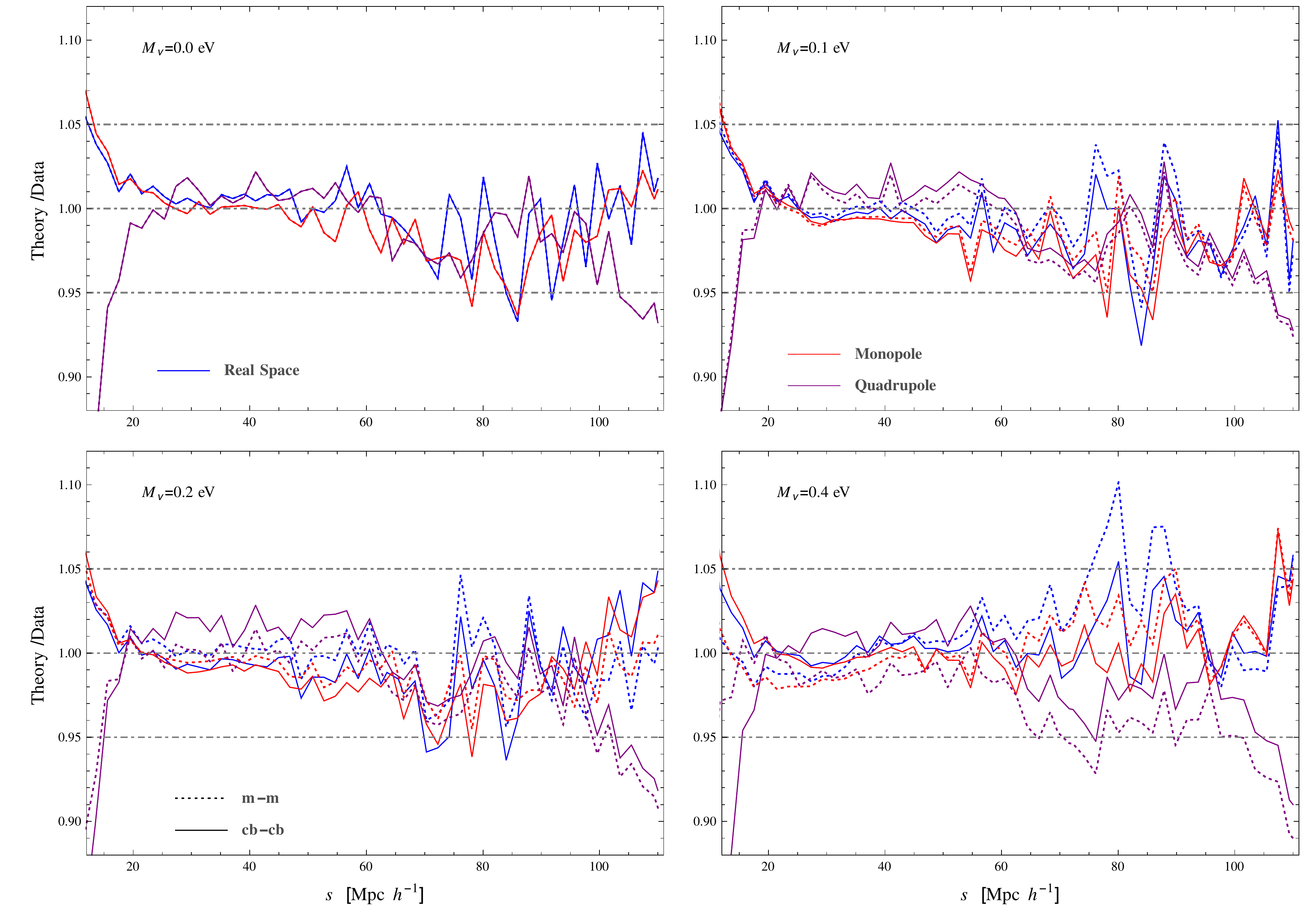}
 	\caption{Ratios of halo correlation functions to simulated data. The different panels show the cases of neutrinos with masses
 	$M_\nu=0,\,0.1,\,0.2,\,0.4$ eV. Solid lines correspond to the
 	biasing scheme applied to the $cb$ fluid and dotted lines to the whole matter fluid. Blue lines are for the real 
 	space correlation function, red lines for the redshift space monopole and purple lines for the quadrupole. 
 	\label{fig:CFhalosRatios}}
 	\end{center}
 \end{figure}

We will now compare our theory for redshift and real space correlation functions to halos obtained from the \textsc{Quijote} simulations. 
The simulation halos are identified using a Friends-of-friends algorithm \cite{1985ApJ...292..371D} run on the CDM particles only, with 
linking length parameter $b=0.2$. The halo mass, therefore is just a sum over the masses of all particles that are associated with an individual halo.  
Here, we consider halos with masses in the range $10^{13.1} M_\odot \,h^{-1}  < M_h  < 10^{13.5} M_\odot \,h^{-1}$.

Each model we test has four free parameters, three biases $b_1$, $b_2$ and $b_{\nabla^2\delta}$ and one EFT parameter $\alpha_\sigma$, that we adjust empirically to fit three simulated data sets: the real space correlation function, and the 
$\ell=0,2$ multipoles of the redshift-space correlation function. We do this for both biasing the $cb$ and $m$ correlation functions. However, since the real space correlation function is not available to real surveys, 
we only use it (for simplicity) to fit to the overall large-scale shift given by $b_1$, and the other three free parameters are estimated by fitting to resdhift-space data directly.

In the presence of massive neutrinos, even linear bias is scale-dependent. In \cite{Castorina:2013wga,Banerjee:2019omr}, it is found that 
\begin{equation}\label{bLS}
 b_\text{LS}(k) = b_c + b_\nu \frac{P_{cb,\nu}(k)}{P_{cb,cb}(k)}  
\end{equation}
is a good approximation for linear bias being the effect of $b_\nu$ more pronounced when biasing the $m$ field, and almost negligible when biasing the $cb$ field, 
because dark matter halos are biased tracers almost entirely of the $cb$ field \cite{Villaescusa-Navarro:2013pva}.
Here, we will expand the ratio of power spectra in powers of $k^2/k^2_\text{FS}$, and obtain an effective bias at large scales
\begin{equation}\label{bLSproxy}
 b_\text{LS}(k) = 1 + b_1 - b_{\nabla^2\delta} k^2   + \cdots.
\end{equation}
That is, we encapsulate the effects of the scale dependent bias as higher-order, curvature biases. This is the main reason why we included it in sections \ref{sect:RSCF} and \ref{sect:zSCF}. However, 
curvature bias serves also to remove large-scale dependencies arising when smoothing the density perturbations \cite{Aviles:2018thp,Schmidt:2012ys}; and furthermore, it is degenerate with counterterms to zero-lag, 
2-point correlators of linear Lagrangian displacements \cite{Vlah:2015sea,Vlah:2016bcl}. Hence, 
these three effects contribute to the estimated value of $b_{\nabla^2\delta}$. Nonetheless, we shall try to keep $b_{\nabla^2\delta}$ consistent with zero as much as possible when biasing the $cb$ field.

Note however that the expansion of the scale-dependent linear bias in powers of $k^2$ of eq.~\eqref{bLSproxy} is formally valid above the free-streaming scale, that may be large. 
For our cosmology at $z=0.5$, this becomes $1/k_\text{FS} \approx 70 \hmpc$ for degenerated neutrinos with total mass $M_\nu =0.1$ eV; for more massive  neutrinos, the free straming scale is smaller. 
In spite of this, we show below that curvature bias provides a good match to the simulated halos, better than if not considered.

\begin{table*}[t]
\centering
\begin{tabular}{lccccc}
\hline
& & & &  \\[-3pt]
$\qquad\qquad$ & $\qquad b_1 \qquad$ & $\qquad b_2 \qquad$ & $\qquad b_{\nabla^2\delta} \qquad$  & $\, \alpha_\sigma \times f_0^2 $   \\ [5pt] 
\hline 
& & & &  \\[-3pt]
(m-m)     & & & &  \\[5pt]
$M_\nu=0.0$ eV      &$0.725$      &$-0.1$   & $0$  & $-7 $ \\[5pt]
$M_\nu=0.1$ eV      &$0.715$       &$-0.2$   & $1$  & $-10$ \\[5pt]
$M_\nu=0.2$ eV      &$0.705$       &$-0.3$   & $1$  & $-12$ \\[5pt]
$M_\nu=0.4$ eV      &$0.67$       &$-0.5$     & $1$  & $-18$ \\[5pt]
 \hline
& & & &  \\[-3pt]
(cb-cb)   & & & &  \\[5pt]
$M_\nu=0.1$ eV      &$0.7$        &$-0.4$   & $0$  & $-14$ \\[5pt]
$M_\nu=0.2$ eV      &$0.69$       &$-0.1$   & $0$  & $-10$ \\[5pt]
$M_\nu=0.4$ eV      &$0.665$       &$-0.2$   & $0.5$  & $-10 $ \\[6pt]
\hline
\end{tabular}
\caption{Lagrangian bias parameters. The top panel shows the parameters when biasing the total matter field, and the lower panel the $cb$ field. 
The units of parameters $ b_{\nabla^2\delta}$ and $ \alpha_\sigma $ are $\text{Mpc}^{2}\,h^{-2}$.
}
\label{tabla:BiasParams}
\end{table*}

Our results are shown in figure \ref{fig:CFhalos}, where we present the correlation functions for the different models, multiplied by factors $r^2$ to cover the whole range of interest. From top to bottom the panels
show the cases $M_\nu=0,\, 0.1,\, 0.2,\, 0.4$ eV. The left column is for the real space correlation function, the middle column for the redshift-space monopole, and the right column for the redshift-space quadrupole. 
The points denote the average of the $100$ different realizations of the simulated data and the error bars capture their scatter. 
The solid lines  are the results of our theory when biasing the $cb$ field, and the dashed lines when biasing the $m$ field. Figure \ref{fig:CFhalosRatios} 
shows the ratios of the LPT  predictions to the $N$-body data points, with blue lines showing the
real space correlation function, red lines the monopole, and purple lines the quadrupole; dashed lines are for the $m$ field and solid lines for the $cb$ field. 
The bias and EFT parameters of these fittings are shown in table \ref{tabla:BiasParams}. We notice that the more commonly used, Eulerian linear bias is related to the linear local Lagrangian bias as $b_1^E=1+b_1$, hence the
biases for the different models are very close to each other, and are slightly smaller for more massive neutrinos. This is a consequence of the almost universal halo mass function for cosmologies 
that have the same $\sigma_8$ \cite{Villaescusa-Navarro:2013pva}. 
All our theoretical results are consistent with the simulations inside the RMS errors down to $r= 20 \hmpc$. 
However, the errors are large, so we performed the fittings trying to match as much as possible the points pondering the large scales, but maintaining a good match inside the error bars at scales $20\hmpc$. 
When we compare between the different models, we note that the $M_\nu=0.4$ eV case performs worse than the other cases, 
particularly when considering biasing with respect to the total matter field (dashed lines of fig.~\ref{fig:CFhalosRatios}).  The bias parameters are degenerate to some extent, especially $b_2$ and $\alpha_\sigma$ for the quadrupole. 
Hence, different combinations of parameters give similar results, here we report those that seem to match the best. However, we found better fits to the $M_\nu=0.4$ eV case by using a negative curvature bias 
$b_{\nabla^2\delta}=-2 \, \text{Mpc}^2 \,h^{-2}$ and a large (also negative) second order local bias $b_2=-1$. But this is unappealing since eqs.~\eqref{bLS} and \eqref{bLSproxy} suggest curvature bias should be positive; moreover, 
a negative curvature bias translates into a positive contribution $k^2 P_L(k)$ to the 1-loop power spectrum.

\end{section}

\begin{section}{Summary and Conclusions}\label{sect:conclusions}

In this work we have developed a PT framework to study the clustering of matter and tracers in cosmologies that contain massive neutrinos. 
A main complication in constructing such theories is that the large neutrino thermal velocities inhibit formation of structures below the free-streaming scale. 
This implies that the linear growth becomes suppressed at scales below it, but behaves similar to a CDM-only Universe at larger scales. This scale-dependent growth is inherited to higher orders in PT, modifying the commonly used EdS kernel for both
Eulerian and Lagrangian treatments. The latter is the subject of this work.

Our Lagrangian theory is presented in Section \ref{sect:EoM}, arriving to the evolution equation for the Lagrangian displacements in eq.~\eqref{eqm}, where the ``additional'' 
free-streaming scale enters through the function $A(k)$.    
In constructing the theory we make use of non-linear mappings of Fourier transforms of functions 
evaluated on Eulerian coordinates to Fourier transforms of the same functions evaluated at Lagrangian positions. These maps between Eulerian and Lagrangian frames have a geometrical origin and introduce new terms into the perturbative expansions, 
ultimately leading to the last two contributions of eq.~\eqref{eqm}, that we refer throughout as ``frame-lagging'', following \cite{Aviles:2017aor}. We obtain the Lagrangian displacements kernels up to third order in PT, 
which are given by eqs.~\eqref{KernelPsi1}, \eqref{KernelPsi2} and \eqref{LPTK3order}. These reduce to the well known EdS kernels for massless neutrinos, as can be shown by simply setting the function $A(k)=3 H^2 /2$, 
and the frame-lagging kernels to zero. However, taking the large scale limit to the LPT kernels show a correct behavior only when the frame-lagging contributions are accounted for, 
since these provide the precise cancellations to reduce the non-linear Lagrangian displacements to those of a pure CDM fluid. Moreover, the second and third order kernels contracted with external wavevectors behave as 
$k^i L_i \propto k^2$ for $k\ll k_\text{FS}$. 

Note that we do not treat neutrinos and CDM on an equal footing. Instead, we choose the Lagrangian displacements to follow the trajectories of the CDM particles only, 
while non-linear neutrinos are modeled as being proportional to the CDM non-linear fluctuations damped by a factor given by the ratio of linear neutrino 
to CDM overdensities; this is similar to what is done in some PT treatments posed in the Eulerian frame \cite{Lesgourgues:2009am,Upadhye:2013ndm}.
We show that the above mentioned approximation for the neutrino overdensities does not yield to UV divergences in loop statistics, as ref.~\cite{Blas:2014hya} claims happens in the Eulerian treatment. 
In our approach the approximation receives contributions from non-linear Lagrangian displacements as given in eq.~\eqref{deltanuProxy}, ensuring a good convergence at large-scales. 
To show more clearly the importance of the frame-lagging, we use the LPT kernels to construct the SPT real space power spectrum, and show that the loop contributions are free of  
unwanted UV divergences  and behave as $k^2 P_L(k)$ for $k\rightarrow 0$, such that the large scales properly decouple from the small scales. 
This is not the case if we do not consider the frame-lagging; instead, in that case, the large-scales receive arbitrary, cut-off dependent contributions from the small scales.  
This small scale sensitivity is a common feature of methods that breaks Galilean invariance or momentum conservation is violated, as in perturbative schemes that approximate the neutrino density by its linear value. 

We use our LPT to construct real and redshift space correlation functions for particles and tracers, using standard tools of CLPT and the Gaussian Streaming Model, with small modifications to account for kernels beyond EdS, already
found in previous works \cite{Aviles:2018saf,Valogiannis:2019nfz}. (Although those works focus on Modified Gravity theories, the expressions for 2-point statistics are valid for general LPT kernels.) 
We compare our analytical results to the \textsc{Quijote} suite of simulations finding a good match inside error bars down to $20 \hmpc$ to the real space and redshift monopole correlation functions of both matter and $cb$ particles with no free parameters. The 
same accuracy is obtained for the redshift-space quadrupole if we include an EFT parameter, as noted in earlier works on the GSM. For halos, we use a simple Lagrangian biasing prescription that includes only density and curvature operators, we found
that this is sufficient to obtain a good agreement to our simulated halos down to $20 \hmpc$ inside the error bars. More complicated biasing schemes can be incorporated if necessary, for example to include tidal bias; as done in \cite{Vlah:2016bcl} for the 
GSM. Our comparisons were performed for biasing the $cb$ fluid and the whole matter fluid, for degenerated massive neutrinos with total mass $M_\nu=0,\,0.1,\,0.2,\,0.4$ eV, all showing the same level of accuracy inside the error bars. When 
comparing to halos we notice that the curvature bias is consistent with zero when the biasing is performed to the $cb$ densities, but not to the $cb+\nu$ fluid. 
This is not surprising, since early works have shown that the linear bias is almost scale-independent for the former case, but not for the latter. 

To our knowledge this work presents the first consistent LPT for CDM clustering in the presence of massive 
neutrinos.\footnote{In ref.~\cite{Peloso:2015jua} the degradation and shift of the BAO peak of the real space correlation function is studied within the 
Lagrangian resummation theory of \cite{Matsubara:2007wj} with the use of EdS kernels.} 
Moreover, this is the first analytical, PT method that accounts for both the effects of RSD and non-linear bias for cosmologies with massive neutrinos. 
Hence, a natural next step is to map our LPT to SPT kernels to obtain the RSD multipoles for the power spectrum. Other interesting avenue of study is the analytical construction 
of marked statistics that up-weights low density regions, as was done in \cite{White:2016yhs,Aviles:2019fli} for MG theories, 
and that recently have been shown to be promising tools for measuring the absolute mass of the neutrinos with surveys data \cite{Massara:2020pli,Philcox:2020fqx}.  

This work also has implications for generating consistent initial conditions for $N$-body simulations in massive neutrino cosmologies with higher order perturbation theory. In $\Lambda$CDM cosmologies, initial conditions are routinely generated using second order Lagrangian Perturbation theory \cite{2012ascl.soft01005C}. Apart from greater accuracy, use of 2LPT also allows for the simulations to be started later, thereby saving valuable computational time. On the other hand, $N$-body simulations with massive neutrinos are generally initialized using the first order Zeldovich Approximation, and therefore need to start at higher redshifts for the same level of accuracy. This can also lead to systematic issues when comparing results from simulations with and without massive neutrinos.
A consistent 2LPT framework in massive neutrino cosmologies alleviates these issues, and building a framework for initializing massive neutrino cosmology simulations with 2LPT initial conditions provides a particularly appealing application of the results derived in this work.

\end{section}

\acknowledgments
We would like to thank Emanuele Castorina, Jorge L. Cervantes-Cota, Martin White and Francisco Villaescusa-Navarro for discussions and suggestions. A.A. acknowledges partial support from Conacyt Grant No. 283151. A.A. acknowledges the KIPAC--PAVES (Program for Astrophysics Visitor Exchange at Stanford) project. AB would like to thank Stanford University and the Stanford Research Computing Center for providing computational resources and support that contributed to these research results. The \textsc{Pylians3}\footnote{https://github.com/franciscovillaescusa/Pylians3} analysis library  was used extensively in this paper.

\appendix

\begin{section}{{\emph k}- and {\emph q}-functions}\label{app:kq} 

In this appendix we show how the functions $A$, $U$ and $W$ appear in the real space correlation function, the pairwise velocity and the pairwise velocity dispersion are
reduced to expressions suitable for numerical integration. We will refer the reader to \cite{Aviles:2017aor,Aviles:2018saf}, and specially to appendix A of \cite{Valogiannis:2019nfz} where all these functions are displayed. These 
articles focus on MG models, but the expressions are valid for generalized kernels. 

We take as example the function $\dot{A}_{ij}(\vq)$, for which we have
\begin{equation}
\dot{A}_{ij}(\vq) = \langle \Delta_i \frac{\dot{\Delta}_j}{f_0 H } \rangle =  \int \Dk{k_1} \Dk{k_2} \big( e^{i\vk_1\cdot \vq_2} -e^{i\vk_1\cdot \vq_1}\big) 
\big( e^{i\vk_2\cdot \vq_2} -e^{i\vk_2\cdot \vq_1}\big) \langle \Psi_i(\vk_1) \frac{\dot{\Psi}_i(\vk_2)}{f_0 H }  \rangle.
\end{equation}

Rotational symmetry and homogeneity imply there are two $|\vk|$-dependent functions, $a(k)$ and $p(k)$, such that
\begin{equation}
\langle \Psi_i(\vk) \frac{\dot{\Psi}_i(\vk')}{f_0 H }  \rangle = (2\pi)^3 \dD(\vk + \vk') \left( a(k)\delta_{ij} + p(k) \frac{k_i k_j}{k^2} \right) = (2\pi)^3 \dD(\vk + \vk') p(k) \frac{k_i k_j}{k^2},   
\end{equation}
where in the last equality we have used the assumption that the Lagrangian displacement is longitudinal, $\Psi_i(\vk) = (\hat{k}_j \Psi_j)\hat{k}_i$. Hence
\begin{equation} \label{AppEqdotAij}
 \dot{A}_{ij}(\vq) = 2 \int \Dk{k} \big(1-e^{i\vk \cdot \vq} \big) \frac{k_ik_j}{k^2} p(k).
\end{equation}
Then, we expand the Lagrangian displacement and its derivative as 
$\Psi=\Psi^{(1)}+\Psi^{(2)} +\cdots$ and  $\dot{\Psi}=\dot{\Psi}^{(1)}+\dot{\Psi}^{(2)} +\cdots$, and obtain that function $p(k)$ is
\begin{equation}
p(k) = \frac{f(k)}{f_0} P^L_{cb}(k) + \frac{9}{49} Q_1^f(k) + \frac{5}{21} \frac{f(k)}{f_0} R_1(k) + \frac{5}{7} R_1^f(k), 
\end{equation}
with functions
\begin{align}
Q_1(k) &= \int \Dk{p} \big[ \Gamma_2(\vk-\vp,\vp)\big]^2 P^L_{cb}(|\vk-\vp|)P^L_{cb}(p), \\
Q_1^f(k) &= \int \Dk{p} \Gamma_2(\vk-\vp,\vp)\Gamma_2^f (\vk-\vp,\vp) P^L_{cb}(|\vk-\vp|)P^L_{cb}(p), \\
R_1(k) &= \int \Dk{p} \frac{21}{10} C_3\Gamma_3(\vk,-\vp,\vp)P^L_{cb}(k)P^L_{cb}(p), \\
R_1^f(k) &=\int \Dk{p} \frac{21}{10} C_3\Gamma_3^f(\vk,-\vp,\vp)P^L_{cb}(k)P^L_{cb}(p).
\end{align}
We have used the ``scalar'' kernels for $k_i \Psi_i$ and $k_i \dot{\Psi}_i$, given by \cite{Valogiannis:2019nfz}
\begin{align}
C_n \Gamma_n(\vk_1,\dots,\vk_n;t) &= k^i_{1\cdots n} L_i^{(n)}(\vk_1\dots,\vk_n;t)  \\
C_n \Gamma_n^f(\vk_1,\dots,\vk_n;t) &= k^i_{1\cdots n} L_i^{f(n)}(\vk_1\dots,\vk_n;t),  
\end{align}
where we choose $C_1=1$ and $C_2=3/7$.
The first order scalar kernels are $\Gamma_1(\vk)=1$ and $\Gamma_1^f(\vk)=f(k)/f_0$. To second order 
\begin{align} \label{Gamma2}
\Gamma_2(\vp_1,\vp_2) &= \left[\mA - \mB \frac{(\vp_1 \cdot \vp_2)^2}{p_1^2 p_2^2}\right]
                       =  \frac{7}{3}\frac{ D^{(2)}(\vp_1,\vp_2) }{ D_+(p_1)D_+(p_2) }, \\     
\Gamma^f_2(\vp_1,\vp_2) 
     &= \Gamma_2(\vp_1,\vp_2) \frac{f(p_1) + f(p_2)}{2 f_0} 
     +  \frac{1}{2f_0 H_0}\left[\dot{\mA} - \dot{\mB} \frac{(\vp_1 \cdot \vp_2)^2}{p_1^2 p_2^2}\right], \nonumber\\
      &=  \frac{1}{2f_0 H}\frac{7}{3}\frac{   \frac{d\,}{d t} D^{(2)}(\vp_1,\vp_2) }{ D_+(p_1)D_+(p_2) },
\end{align}
where $\mA,\mB=\mA,\mB(\vp_1,\vp_2)$. The third order scalar kernels are
\begin{align}
C_3 \Gamma_3(\vp_1,\vp_2,\vp_3) &= \frac{D_+^{(3)s}(\vp_1,\vp_2,\vp_3)}{D_+(\vp_1)D_+(\vp_2)D_+(\vp_3)},  \\
C_3 \Gamma^f_3(\vp_1,\vp_2,\vp_3) &= \frac{1}{3f_0H} \frac{\frac{d\,}{d t} D_+^{(3)s}(\vp_1,\vp_2,\vp_3)}{D_+(\vp_1)D_+(\vp_2)D_+(\vp_3)}.
\end{align}

Now, with the solid angle integral identity
\begin{equation}
 \frac{1}{4\pi} \int d \Omega_{\hat{\vk}} e^{i \vk \cdot \vq} \hat{k}_i \hat{k}_j = \frac{j_1(k q)}{k q} \delta_{ij} - j_2(k q) \hat{q}_i \hat{q}_j
\end{equation}
we can bring eq.~\eqref{AppEqdotAij} to
\begin{equation}
 \dot{A}_{ij}(\vq) = \dot{X}(q) \delta_{ij} + \dot{Y}(q) \hat{q}_i \hat{q}_j,
\end{equation}
with
\begin{align}
 \dot{X}(q) &= \frac{1}{\pi^2} \int dk \, p(k) \left[ \frac{1}{3} - \frac{j_1(k q)}{k q} \right], \\
 \dot{Y}(q) &= \frac{1}{\pi^2} \int dk \, p(k)  j_2(k q).
\end{align}

We notice that for the massless neutrino case, the scalar kernels reduce to
\begin{equation}
 \Gamma_n^f \simeq  \Gamma_n,  \qquad \text{($M_\nu=0$)},
\end{equation}
which further implies that functions $Q^f$ and $R^f$ reduce to $Q$ and $R$, and $\dot{A}_{ij}(\vq)$ to the standard result in $\Lambda$CDM (see \cite{Wang:2013hwa}).

Using the same methods presented here, one can obtain all the ``undotted'' and ``dotted'' functions $A$, $W$ and $U$, necessary to construct the correlation functions in CLPT and the GSM. 
All these functions are displayed in appendix A of ref.~\cite{Valogiannis:2019nfz}, which are valid for general kernels $\Gamma$ and $\Gamma^f$. In that reference one can find also how to introduce
tidal bias into the GSM for generalized LPT kernels.

\end{section}

 \bibliographystyle{JHEP}  
 \bibliography{bib.bib}

\end{document}